\documentclass[12pt]{article} 
\usepackage{latexsym}
\usepackage{amsmath}
\usepackage{pstricks}
\usepackage{indentfirst}
\usepackage{amsthm}
\newcommand{\ad}{^\dagger }
\newcommand{\becs}{\begin{cases}}
\newcommand{\bem}{\begin{matrix}}
\newcommand{\besp}{\begin{split}}

\newcommand{\colo}{\,\hbox{:}\,}

\newcommand{\encs}{\end{cases}}
\newcommand{\enm}{\end{matrix}}
\newcommand{\ensp}{\end{split}}

\newcommand{\lgl}{\langle }

\newcommand{\ot}{\otimes }

\newcommand{\ra}{\rightarrow }

\newcommand{\rgl}{\rangle }

\newcommand{\Tr}{{\rm Tr}}

\newcommand{\scriptl}[1] {{\cal #1}}

\newcommand{\AS}{\scriptl A }
\newcommand{\BS}{\scriptl B }

\newcommand{\HS}{\scriptl H }

\newcommand{\PS}{\scriptl P }

\newcommand{\ST}{\scriptl S } 

\newcommand{\VS}{\scriptl V }

\newcommand{\XS}{\scriptl X }

\newcommand{\al}{\alpha }
\newcommand{\bt}{\beta }
\newcommand{\gm}{\gamma }
\newcommand{\Gm}{\Gamma }
\newcommand{\dl}{\delta }

\newcommand{\zt}{\zeta }

\newcommand{\kp}{\kappa }

\newcommand{\Lm}{\Lambda }
\newcommand{\sg}{\sigma }

\theoremstyle{definition} 
\theoremstyle{plain} 

\newtheoremstyle{break}%
{}{}%
{}
{}
{\itshape}
{.}
{ }
{}

\theoremstyle{break}
\newtheorem{theorem}{Theorem}

\begin{document}

\title{Channel kets, entangled states, and the location of quantum information}

\author{Robert B. Griffiths
\thanks{Electronic mail: rgrif@cmu.edu}\\ 
Department of Physics,
Carnegie-Mellon University,\\
Pittsburgh, PA 15213, USA}


\date{Version of 3 Feb. 2005}

\maketitle  

\begin{abstract}
	
The well-known duality relating entangled states and noisy quantum channels is
expressed in terms of a \emph{channel ket}, a pure state on a suitable
tripartite system, which functions as a pre-probability allowing the
calculation of statistical correlations between, for example, the entrance and
exit of a channel, once a framework has been chosen so as to allow a consistent
set of probabilities.  In each framework the standard notions of ordinary
(classical) information theory apply, and it makes sense to ask whether
information of a particular sort about one system is or is not present in
another system.  Quantum effects arise when a single pre-probability is used to
compute statistical correlations in different incompatible frameworks, and
various constraints on the presence and absence of different kinds of
information are expressed in a set of all-or-nothing theorems which generalize
or give a precise meaning to the concept of ``no-cloning.'' These theorems are
used to discuss: the location of information in quantum channels modeled using
a mixed-state environment; the  CQ  (classical-quantum) channels introduced by
Holevo; and the location of information in the physical carriers of a quantum
code.  It is proposed that both channel and entanglement problems be classified
in terms of pure states (functioning as pre-probabilities) on systems of $p\geq
2$ parts, with mixed bipartite entanglement and simple noisy channels belonging
to the category $p=3$, a five-qubit code to the category $p=6$, etc.; then by
the dimensions of the Hilbert spaces of the component parts, along with other
criteria yet to be determined.
\end{abstract}


	\section{Introduction}
\label{sct1}

Understanding entangled states and the properties of quantum channels are two
central issues in quantum information theory.  At least in a formal sense they
are the same problem: the duality mapping one into the other has been discussed
explicitly in recent work \cite{Kyl02,VrVr03,ZcBn04,ArPt04}, and employed for
various purposes in a much larger collection of papers; see
\cite{CDKL01,RmWr03,HrSR03,Rska03,Hmda03,DCHr04} for a few examples in addition
to those in the extensive bibliography in \cite{ZcBn04}.  The early work most
often cited is \cite{Jmlk72,Choi75}, though the basic idea is not complicated,
and has undoubtedly been rediscovered many times.  Nonetheless, one has the
impression that this duality has yet to be fully exploited, and much more could
be done to relate the concepts used in discussing entanglement, and the large
number of proposed measures of entanglement, to the ideas employed for thinking
about quantum channels, and the definitions of many different sorts of channel
capacity.  Perhaps a barrier to its full utilization is the fact that this
duality remains something of a mathematical abstraction whose connection with
more physical ideas has not been totally clear.
One aim of the present paper is to relate this duality to concepts of quantum
\emph{information}.  To be sure, ``information'' as it applies to the quantum
domain is not at present a very precise concept; the appropriate definitions
remain the subject of current research and occasional controversy
\cite{BrZl99,BrZl01,DtHy00,Grff02,Tmps03,Dwll03}.  The term is used here in the
very broad sense of \emph{statistical correlation}, an idea familiar in
classical physics and classical information theory, which deserves to be better
understood and more widely applied in the quantum domain.

The duality under discussion can be formulated in various ways.  One which
seems particularly helpful characterizes a noisy quantum channel using a
\emph{channel ket}, an entangled pure state on a suitable tripartite system;
see Sec.~\ref{sct2c} for the precise definition.  While this idea is (at least)
implicit in previous work, the main emphasis has been on the duality between a
density operator describing a mixed state of a bipartite system and what we
here call a \emph{dynamical operator} (following \cite{ZcBn04}, where the term
\emph{dynamical matrix} is used), closely connected to the superoperator
describing the action of a quantum channel.  The channel ket is obtained by
``purifying'' the dynamical operator using a (possibly fictitious) reference
system; in turn, the dynamical operator is a partial trace over the projector
corresponding to the channel ket.  This relationship is well known and
frequently exploited in the case of mixed entangled states (see, e.g., p.~110
of \cite{NlCh00}).  What is less well known is that there are certain
advantages, both formal and conceptual, in using pure states rather than (or at
least in addition to) mixed states when discussing the \emph{location} of
quantum information --- see Sec.~\ref{sct4} --- and thus occasions when a
channel ket provides insights not directly available from a dynamical operator.
It should be noted that the principal role of a channel ket is the same as that
of a dynamical operator or a density operator: it allows one to calculate
probabilities for various properties of a quantum system.  These probabilities
determine the statistical correlations between events at different times that
provide a physical description of a quantum channel, just as the statistical
correlations between separate quantum systems at a given moment of time provide
a physical description of entanglement.

The remainder of this paper is structured in the following way.  After
introducing some conventions on notation in Sec.~\ref{sct2a}, the basic map-ket
duality is reviewed in Sec.~\ref{sct2b}; our treatment differs from previous
ones mainly in maintaining what we think is a helpful distinction between
operators and their matrices.  Channel kets are defined in Sec.~\ref{sct2c},
with some simple examples in Sec.~\ref{sct2d}.  Brief remarks on the
inverse problem of turning entangled states into channels are found in
Sec.~\ref{sct2e}.

Quantum information in the sense of statistical correlations is the topic of
Sec.~\ref{sct3}.  Sample spaces and probabilities for quantum systems are
discussed in Sec.~\ref{sct3a}, and applied to correlated systems in
\ref{sct3b}.  The notion of \emph{particular types} of information about
certain subsystems being present or absent in other subsystems, which is
central to our later discussions, is introduced in Sec.~\ref{sct3c} for
entangled states, and extended to quantum channels, where the ideas are very
similar modulo a partial transpose, in Sec.~\ref{sct3d}.  These definitions are
\emph{qualitative} and do not depend upon any quantitative measures of
information.  We believe, however, that once correlations have been defined in
a consistent manner, there is no barrier to using quantitative information
measures, such as Shannon's mutual information; this should take care of
the objections raised in \cite{BrZl01}.  The point of view adopted here is
consistent with and an extension of that in \cite{Grff02}.

Following this, Sec.~\ref{sct4} contains a set of ``all or nothing'' theorems
that apply to qualitative aspects of information.  These theorems have a
number of interesting consequences, some of which are discussed in
Sec.~\ref{sct5}, where they are applied to two special types of quantum
channels --- mixed-state environment and ``CQ'' channels --- and to the problem
of the location of information in quantum codes.  In Sec.~\ref{sct6} we propose
a scheme, at present rather tentative, for classifying both entanglement and
channel problems in terms of pure-state entanglement on $p$-part systems.

	The conclusion, Sec.~\ref{sct7}, has both a summary and a list of open
problems. Appendix~\ref{scta} contains the proofs of the theorems of
Sec.~\ref{sct4}, and App.~\ref{sctb} a particular result on bipartite entangled
kets used in App.~\ref{scta}.

	\section{Map-Ket Duality and Channel Kets}
\label{sct2}

	\subsection{Notation}
\label{sct2a}
 
We shall use subscripts $a,b,c$, etc., and sometimes numbers, to label
different subsystems of a system with several parts.  The Hilbert space $\HS_a$
is associated with system $\ST_a$, the tensor product
\begin{equation}
  \HS_{ab} = \HS_a\ot\HS_b
\label{eqn1}
\end{equation}
with the combined system $\ST_{ab}$ consisting of $\ST_a$ and $\ST_b$, and so
forth.  For a ket $|\psi\rgl\in\HS_{abc}$ we use the notation
\begin{equation}
  \psi = [\psi] = |\psi\rgl\lgl\psi|,
\label{eqn2}
\end{equation}
where the square brackets distinguish a dyad from other types of operator.
Partial traces are denoted by 
\begin{equation}
  \psi_{ab} = \Tr_c(\psi),\quad \psi_a = \Tr_b(\psi_{ab}) = \Tr_{bc}(\psi),
\label{eqn3}
\end{equation}
and so forth, both for dyads and other operators. Operators on the Hilbert
space $\HS_a$ themselves form a Hilbert space $\hat\HS_a$, with inner product
$\lgl A,A'\rgl = \Tr(A\ad A')$.

Because the subscript position is used to label the (sub)system, indices are
often written as superscripts in circumstances in which they are not likely to
be confused with exponents.  Thus $\PS=\{|p^j\rgl\}$ denotes an
\emph{orthonormal basis} for the Hilbert space $\HS_p$ of dimension $d_p$, with
$j$ taking values between $0$ and $d_p-1$.  Two such bases $\PS$ and 
$\bar\PS = \{|\bar p^j\rgl\}$ are called \emph{mutually unbiased} if
\begin{equation}
  |\lgl \bar p^j|p^k\rgl| = 1/\sqrt{d_p},
\label{eqn4}
\end{equation}
independent of $j$ and $k$.

More generally, we shall be
interested in a \emph{projective decomposition of the identity} of $\HS_p$,
hereafter called a ``decomposition'', a collection $\{P^k\}$ of projectors
summing to the identity $I_p$ and mutually orthogonal to each other,
\begin{equation}
  I_p = \sum_k P^k,\quad P^k P^l = \dl_{kl} P^k.
\label{eqn5}
\end{equation}
(Recall that a projector is a Hermitian operator equal to its square, so its
eigenvalues are 0 and 1.)  No confusion arises if the same symbol $\PS$ is used
to denote an orthonormal basis $\{|p^j\rgl\}$ or the collection $\{[p^j]\}$ of
the corresponding projectors.

Given an orthonormal basis $\{|a^j\rgl\}$ of $\HS_a$, any ket $|\psi\rgl$
in $\HS_{ab}$ can be expanded in the form
\begin{equation}
  |\psi\rgl = \sum_j |a^j\rgl\ot |\bt^j\rgl,
\label{eqn6}
\end{equation}
where $|\bt^j\rgl=\lgl a^j|\psi\rgl$ is uniquely determined by $|\psi\rgl$ and
$|a^j\rgl$.  If the $\{|\bt^j\rgl\}$ are mutually orthogonal, we shall call
\eqref{eqn6} a \emph{Schmidt} expansion, and sometimes write it in the
alternative form 
\begin{equation}
  |\psi\rgl = \sum_j \sqrt{p_j}\, |a^j\rgl\ot |b^j\rgl,
\label{eqn7}
\end{equation}
with the $\{|b^j\rgl\}$ an orthonormal basis of $\HS_b$, and the
$p_j$ summing to 1 when $|\psi\rgl$ is normalized, $\lgl\psi|\psi\rgl=1$. 
By the \emph{support} of an operator $A$ we shall mean the smallest 
projector $P$ such that 
\begin{equation}
  PAP=A,
\label{eqn8}
\end{equation}
or the subspace $\PS$ onto which this $P$ projects.  The $rank$ of $A$
is the trace of $P$, or the dimension of $\PS$, or the number of
nonzero (positive) eigenvalues of $A\ad A$, or the rank of the matrix
representing $A$. 

	\subsection{Maps and kets}
\label{sct2b}

Given any linear map $M:\HS_a\ra\HS_b$ and an orthonormal basis
$\AS=\{|a^j\rgl\}$ of $\HS_a$, one can define a corresponding ket
\begin{equation}
  |\psi\rgl = \sum_j |a_j\rgl \ot M|a_j\rgl
\label{eqn9}
\end{equation}
on the tensor product $\HS_{ab}$.  Conversely, given such a ket, one can always
expand it in the form \eqref{eqn6} using the basis $\AS$, and define
a map $M$ by 
\begin{equation}
  M|a^j\rgl = |\bt^j\rgl,
\label{eqn10}
\end{equation}
and its extension to all of $\HS_a$ by linearity.  These two formulas define
the \emph{map-ket duality} used throughout the rest of this paper.

The duality depends, obviously, on the choice of orthonormal basis $\AS$; given
a different choice $\bar\AS=\{|\bar a^j\rgl\}$, a given map will lead to a
different ket, and vice versa.  For those who (like the author) prefer to write
formulas whenever possible in basis-independent form, this dependence is
somewhat annoying.  One can get around it, as in \cite{VrVr03,ZcBn04}, by
always using a single basis.  We prefer to maintain the usual distinction
between operators and matrices.  The price for doing this is not exorbitant,
because the basis dependence can always be expressed in terms of a suitable
unitary transformation on $\HS_a$.  And if one is primarily concerned with
concepts which are \emph{invariant under local unitaries}, meaning unitary
operations which are tensor products of unitaries on individual subsystems,
such basis dependence is not intolerable.

\begin{figure}[h]
$$
\begin{pspicture}(-1,-0.3)(4.0,1.3) 
\def\lwd{0.035} 
\psset{
labelsep=2.0,
arrowsize=0.150 1,linewidth=\lwd}
\def\rectc(#1,#2){%
\psframe[fillcolor=white,fillstyle=solid](-#1,-#2)(#1,#2)}
\def\squ{\rectc(0.35,0.35)}
\def\vvv#1{\vrule height #1 cm depth #1 cm width 0pt}
\def\rbrac{$\left.\vvv{0.5}\right\}$}
\def\lbrac{$\left\{\vvv{0.5}\right.$}
\psline{->}(0.0,1.0)(3.0,1.0)
\psline{->}(0.0,0.0)(3.0,0.0)
\rput(1.5,0){\squ}\rput(1.5,0.0){$M$}
\rput(-0.1,0.5){\lbrac}
\rput(3.1,0.5){\rbrac}
\rput[b](0.5,1.1){$a$}
\rput[b](0.5,0.1){$v$}
\rput[b](2.5,1.1){$a$}
\rput[b](2.5,0.1){$b$}
\rput[r](-0.3,0.5){$|\phi\rgl$}
\rput[l](3.3,0.5){$|\psi\rgl$}
\end{pspicture}
$$
\caption{%
Circuit illustrating map $M$ - ket $|\psi\rgl$ duality.}
\label{fgr1}
\end{figure}
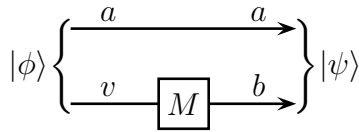

A way of visualizing the relationship between $|\psi\rgl$ and $M$, and for
understanding the ambiguity associated with the choice of basis, is indicated
by the circuit in Fig.~\ref{fgr1}, where $|\phi\rgl$ is a fully-entangled state
\begin{equation}
  |\phi\rgl = \sum_j |a^j\rgl\ot|v^j\rgl
\label{eqn11}
\end{equation}
on the system $\HS_a\ot\HS_v$,  $\HS_v$ is an auxiliary Hilbert space of the
same dimension of $\HS_a$, and $M|v^j\rgl=|\bt^j\rgl$, as in \eqref{eqn10}.
Choosing a different fully-entangled state in place of \eqref{eqn11} would
result in a different relationship between $M$ and $|\psi\rgl$; this is
precisely the ambiguity previously discussed, and provides a good way of
analyzing it.

\subsection{Channel kets and superoperators}
\label{sct2c}

\begin{figure}[h]
$$
\begin{pspicture}(-4.6,-0.2)(3.7,1.2) 
\def\rectc(#1,#2){%
\psframe[fillcolor=white,fillstyle=solid](-#1,-#2)(#1,#2)}
	\def\pta{
\psline{->}(0.0,1.0)(3.0,1.0)
\psline{->}(0.0,0.0)(3.0,0.0)
\rput(1.5,0.5){\rectc(0.7,0.6)}
\rput(1.5,0.5){$T$}
\rput[r](-0.1,1.0){$|a\rgl$} 
\rput[l](3.1,1.0){$b$}
\rput[r](-0.1,0.0){$|e^0\rgl$}
\rput[l](3.1,0.0){$f$}
	}
	\def\ptb{
\psline{->}(0.0,1.0)(3.0,1.0)
\psline{->}(1.5,0.0)(3.0,0.0)
\rput(1.5,0.5){\rectc(0.7,0.6)}
\rput(1.5,0.5){$V$}
\rput[r](-0.1,1.0){$|a\rgl$} 
\rput[l](3.1,1.0){$b$}
\rput[l](3.1,0.0){$f$}
	}
\rput(-4.0,0){\pta}
\rput(-0.4,0.5){$=$}
\rput(0.3,0){\ptb}
\end{pspicture}
$$
\caption{%
Quantum channel using a unitary transformation $T$ or isometry $V$.}
\label{fgr2}
\end{figure}
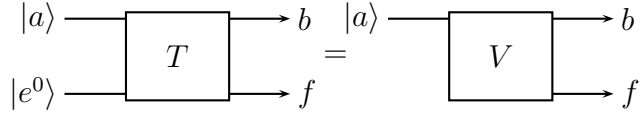

We adopt the following by now fairly standard model for a noisy quantum
channel.  A unitary time transformation $T$ maps the tensor product $\HS_{ae}$
of the Hilbert space $\HS_a$ of the \emph{channel entrance} $\ST_a$ and the
space $\HS_e$ of the (initial) \emph{environment} $\ST_e$, at some initial time
to $\HS_{bf}=\HS_b\ot\HS_f$, corresponding to the \emph{channel exit} or
\emph{output} $\ST_b$ and \emph{environment} $\ST_f$, at some later time,
Fig.~\ref{fgr2}.  Initially the environment is in a \emph{fixed} pure state
$|e^0\rgl$, whereas the initial state of the channel is arbitrary, not fixed in
advance.  Because $|e^0\rgl$ is fixed, the only relevant effect of the unitary
operator $T$ is that embodied in the isometry $V:\HS_a\ra\HS_{bf}$ defined by
\begin{equation}
  V|a\rgl = T\left(|a\rgl\ot|e^0\rgl\right),
\label{eqn12}
\end{equation}
and shown schematically in the second part of Fig.~\ref{fgr2}.  Often $\HS_a$
and $\HS_b$ are identified with each other, and $\HS_e$ with $\HS_f$.
Maintaining the distinction both allows for the possibility, sometimes useful,
that the dimensions of $\HS_a$ and $\HS_b$ may be different, but equally
important permits a distinct label.  It is sometimes useful to assume that
the environment is initially in a mixed, rather than a pure state, see
Sec.~\ref{sct5a}, but there is no loss in generality in assuming a pure state
$|e^0\rgl$, since a mixed state can always be purified by introducing an
auxiliary system, which can then be thought of as part of $\ST_e$.

The \emph{channel ket} $|\Psi\rgl\,\in\HS_{abf}$ is defined as the ket dual to
$V$ in the sense of Sec.~\ref{sct2b},
\begin{equation}
  \sqrt{d_a}\, |\Psi\rgl = \sum_j |a^j\rgl\ot V |a^j\rgl 
  \,\in \HS_{a}\ot\HS_{bf},
\label{eqn13}
\end{equation}
using an orthonormal basis $\AS=\{|a^j\rgl\}$ of $\HS_a$.  The normalization
$\|\Psi\|=1$ is of no great importance --- which is why $\sqrt{d_a}$ is placed
on the left side of this equation --- but does simplify certain formulas.
Notice that $|\Psi\rgl$ is a pure state on a tripartite system.

\begin{figure}[h]
$$
\begin{pspicture}(-1.0,-0.2)(4.0,2.2) 
\def\rectc(#1,#2){%
\psframe[fillcolor=white,fillstyle=solid](-#1,-#2)(#1,#2)}
\def\vertdash(#1){\psline[linestyle=dashed,linewidth=0.01](0.0,0.0)(0.0,#1)}
\def\vtpair{\vertdash(1.0)\rput(0.1,0){\vertdash(1.0)}%
\psline(0.0,0.0)(0.1,0.0)\psline(0.0,1.0)(0.1,1.0)}
\def\vvv#1{\vrule height #1 cm depth #1 cm width 0pt}
\def\rbrac{$\left.\vvv{1.0}\right\}$}
\def\lbrac{$\left\{\vvv{0.5}\right.$}
\psline{>->}(0.0,2.0)(3.0,2.0)
\psline{>->}(0.0,1.0)(3.0,1.0)
\psline{->}(1.5,0.0)(3.0,0.0)
\rput(1.5,0.5){\rectc(0.7,0.6)}
\rput(1.5,0.5){$V$}
\rput(-0.5,1.5){\lbrac}
\rput[r](-0.1,2.0){$a$} 
\rput[r](-0.1,1.0){$v$} 
\rput[r](-0.7,1.5){$|\phi\rgl$}
\rput[l](3.1,2.0){$a$}
\rput[l](3.1,1.0){$b$}
\rput[l](3.1,0.0){$f$}
\rput[l](3.3,1.0){\rbrac}
\rput[l](3.7,1.0){$|\Psi\rgl$}
\end{pspicture}
$$
\caption{%
Circuit for visualizing the channel ket $|\Psi\rgl$.}
\label{fgr3}
\end{figure}
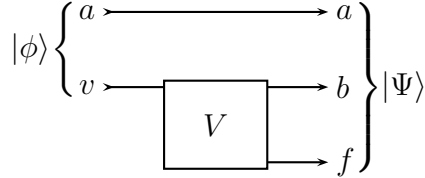

The channel ket can be visualized using Fig.~\ref{fgr3}, the obvious analog of
Fig.~\ref{fgr1}, as obtained by transmitting the $\ST_v$ part of the
fully-entangled state \eqref{eqn11} through the channel, while preserving the
$\ST_a$ part unchanged.  It is important to distinguish the \emph{definition}
of the channel ket, given in \eqref{eqn13}, from this visualization, in that
$|\Psi\rgl$ is a mathematical object which functions as a pre-probability, used
to calculate probabilities of various events or processes associated with the
channel, as discussed in Sec.~\ref{sct3}, quite apart from whether the channel
is being used in the manner just described.

Following the notation of Sec.~\ref{sct2a}, the symbol $\Psi$ denotes the dyad
$|\Psi\rgl\lgl\Psi|$, and subscripts are used to indicate its partial traces. 
Of particular importance is the \emph{dynamical operator}
\begin{equation}
  R := \Psi_{ab} =\Tr_f(\Psi)\,\in \hat\HS_{ab},
\label{eqn14}
\end{equation}
which corresponds to the dynamical matrix defined in \cite{ZcBn04} (apart 
from the order $ab$ as against $ba$); the latter
is $R$ with a particular choice of basis.  Since $\Psi$ is a positive operator,
so is $R$, and given the normalization in \eqref{eqn13}, $R$ has unit trace.
In addition, because $V$ is an isometry, 
\begin{equation}
  R_a = \Tr_b(R) = \Psi_a = I_a/d_a. 
\label{eqn15}
\end{equation}
Thus $R$ is a density operator for the bipartite system $\HS_{ab}$, with the
special property that $R_a$ is proportional to the identity.  Hence whatever
intuition one possesses for mixed states on bipartite systems can at once be
applied to $R$; e.g., one can ask if it is separable, and if not, how entangled
it is according to any of the numerous measures of mixed-state entanglement,
etc. 

But in addition, $R$ completely determines the properties of the noisy quantum
channel, that is, the channel superoperator, up to a unitary transformation of
the channel input $\HS_a$ corresponding to different choices for the basis used
in the definition \eqref{eqn13}.  The channel superoperator $\VS$ is the map
from $\hat\HS_a$ to $\hat\HS_b$ defined by
\begin{equation}
  \VS(A) = \Tr_f(VAV\ad)
\label{eqn16}
\end{equation}
for any operator $A$ in $\hat\HS_a$.  To explore how $\VS$ is related to 
$R$, it is helpful to choose an orthonormal basis $\{|f^l\rgl\}$ for
$\HS_f$, and expand $|\Psi\rgl$ as
\begin{equation}
  |\Psi\rgl = \sum_l |\kp^l\rgl \ot |f^l\rgl \,\in \HS_{ab}\ot\HS_f.
\label{eqn17}
\end{equation}
We shall refer to the expansion coefficients $\{|\kp^l\rgl\}$
as \emph{Kraus kets}, in that they can, using the duality introduced in
Sec.~\ref{sct2b}, be turned into maps
\begin{equation}
  \sqrt{d_a}\, |\kp^l\rgl = \sum_j |a^j\rgl\ot K_l |a^j\rgl,
\label{eqn18}
\end{equation}
where the $K_l$ are the usual Kraus operators, labeled by subscripts as is the
usual convention.  They can be used to express the channel superoperator in the
familiar form
\begin{equation}
  \VS(A)=\sum_l K_l A K\ad_l.
\label{eqn19}
\end{equation}
The usual normalization $\sum_l K_l\ad K_l=I_a$ is the counterpart of
\eqref{eqn15}.

The $K_l$ no longer depend upon the arbitrary choice of basis $\{|a^j\rgl\}$
used in defining $|\Psi\rgl$, as this dependence is undone when kets are
changed to maps (using the same basis) in \eqref{eqn18}, but they do depend
upon the choice of basis $\{|f^l\rgl\}$.  One can eliminate, or (in degenerate
cases) at least mitigate this arbitrariness by making \eqref{eqn17} a Schmidt
expansion, so that the $\{|\kp^l\rgl\}$ are orthogonal to one another or,
equivalently,
\begin{equation}
  \Tr_a(K\ad_l K_m)= 0 \text{ for } l\neq m. 
\label{eqn20}
\end{equation}
In that case the number of nonzero terms in \eqref{eqn17}, what could be called
the \emph{Kraus rank} of the channel superoperator, is the rank (in the
ordinary sense) of the dynamical operator $R=\Psi_{ab}$. 

Combining \eqref{eqn17} and \eqref{eqn18}, one obtains the expression
\begin{equation}
  R = \Psi_{ab} = \sum_l [\kp^l] = \frac{1}{d_a}
 \sum_{j,k} |a^j\rgl\lgl a^k|\ot \sum_l K_l|a^j\rgl\lgl a^k|K\ad_l,
\label{eqn21}
\end{equation}
for $R$, and 
from it another formula
\begin{equation}
  \VS(A) = \Tr_a[(A\ot I)Q],
\label{eqn22}
\end{equation}
for the channel superoperator in terms of the \emph{transition operator} $Q$,
the partial transpose
\begin{equation}
  Q=R^{T\AS} =  \frac{1}{d_a}
 \sum_{j,k} |a^k\rgl\lgl a^j|\ot \sum_l K_l|a^j\rgl\lgl a^k|K\ad_l\,
  \in\hat\HS_{ab}
\label{eqn23}
\end{equation}
 of the dynamical operator with respect to the basis $\AS=\{|a^j\rgl\}$.  Once
again, by using this same basis a second time, its effect in defining
$|\Psi\rgl$ has been undone, and $Q$ is independent of the basis, consistent
with the fact that the superoperator $\VS$ in \eqref{eqn22} also does not
depend upon the choice of basis.  Despite their close relationship, $Q$ and $R$
are very different types of operators; the latter is positive, and the former,
while it is Hermitian, will typically have negative as well as positive
eigenvalues.

The superoperator $\VS$ is a map from $\hat\HS_a$ to $\hat\HS_b$, so it can be
represented as a matrix once orthonormal operator bases have been defined for
these two spaces.  There are many ways of choosing such bases, but one that is
particularly convenient when $\HS_a$ and $\HS_b$ are qubits is the \emph{Pauli
representation} using $\{\sg_a^j\}$, with $j=0$ the identity and $j=1,2,3$ the
$x,y$, and $z$ Pauli matrices in the standard basis of $\HS_a$, and similarly
$\{\sg_b^j\}$.  Expanding the transition operator $Q$ in the Pauli form --- see
the examples in Sec.~\ref{sct2d} --- often provides a clearer notion of what a
noisy channel ``does'' than is evident by looking at the Kraus operators. There
are various ways of generalizing this representation to higher-dimensional
spaces.  For the case of a channel superoperator there is some advantage to
using a basis of Hermitian operators, rather than unitaries as in
\cite{Sctt04}, because the resulting matrix is real.  If the basis is again
denoted by $\{\sg^j\}$, with $0\leq j \leq d^2-1$ for a $d$-dimensional Hilbert
space, one can again let $\sg^0$ be the identity, so that the orthogonality
condition
\begin{equation}
  \Tr(\sg^j\sg^k) =  \dl_{jk}d
\label{eqn24}
\end{equation}
implies that $\sg^j$ for $j>0$ has zero trace --- this makes it easy to take
partial traces of operators written in Pauli form.

\subsection{Examples of one qubit channels}
\label{sct2d}

We use the names for one qubit channels employed in Sec.~8.3 of \cite{NlCh00},
but employ $p$ in a way which identifies it as the probability of an error.
The channel kets are sums of terms of the form $|ab\rgl\ot|f\rgl$, where $a$
and $b$ are either 0 or 1, but $f$ sometimes takes larger values.

The bit flip channel is described by
\begin{equation}
  \sqrt{2}\,|\Psi\rgl=\sqrt{1-p}\,\Bigl( |00\rgl + |11\rgl\Bigr)\ot |0\rgl +
  \sqrt{p}\,\Bigl( |01\rgl + |10\rgl\Bigr)\ot |1\rgl
\label{eqn25}
\end{equation}
leading to a transition operator
\begin{equation}
  4Q = I +\sg_a^1\sg_b^1 + (1-2p)[\sg_a^2\sg_b^2 +\sg_a^3\sg_b^3]
\label{eqn26}
\end{equation}
in the Pauli representation.
The dynamical operator $R$ is the same except for a minus sign multiplying
the term $\sg_a^2\sg_b^2$, reflecting the fact that $\sg^y$ changes sign
when transposed. 
   
For the amplitude damping channel the corresponding expressions are
\begin{equation}
  \sqrt{2}\,|\Psi\rgl=\sqrt{1-p}\,\Bigl( |00\rgl + |11\rgl\Bigr)\ot |0\rgl +
  \sqrt{p}\,|10\rgl\ot |1\rgl,
\label{eqn27}
\end{equation}
\begin{equation}
  4Q = I +p\sg_b^3 +\sqrt{1-p}\,\Bigl(\sg_a^1\sg_b^1 + \sg_a^2\sg_b^2\Bigr)
  +(1-p)\sg_a^3\sg_b^3.
\label{eqn28}
\end{equation}
A depolarizing channel requires a larger environment:
\begin{align}
  2|\Psi\rgl &=
 \sqrt{2-3p}\,\Bigl( |00\rgl + |11\rgl\Bigr)\ot |0\rgl
  +\sqrt{p}\,\Bigl( |00\rgl - |11\rgl\Bigr)\ot |1\rgl
\notag\\
  &+\sqrt{2p}\,\Bigl( |01\rgl\ot|2\rgl + |10\rgl\ot|3\rgl\Bigr),
\label{eqn29}
\end{align}
\begin{equation}
  4Q = I +(1-2p)\Bigl(\sg_a^1\sg_b^1 + \sg_a^2\sg_b^2 + \sg_a^3\sg_b^3\Bigr).
\label{eqn30}
\end{equation}
Again, the dynamical operator $R$ is obtained by changing the sign of
$\sg_a^2\sg_b^2$.

\subsection{From entangled states to channels}
\label{sct2e}

As shown in Sec.~\ref{sct2c}, any noisy channel modeled as in Fig.~\ref{fgr2}
can be mapped onto an equivalent entangled ket $|\Psi\rgl$ on a tripartite
system, and thence onto a density operator whose partial transpose determines
the channel superoperator.  Can one do the reverse, starting with a tripartite
ket $|\Psi\rgl$ or a bipartite density operator $R$?  Yes, aside from the
condition that $\Psi_a$ (or $R_a$) be proportional to the identity operator
$I_a$.  But if this is not true, can one still turn an entanglement problem
into a channel problem?  There are at least two approaches, each with
advantages and disadvantages.

The first is to begin with a unitary operator or isometry as in
Fig.~\ref{fgr2}, but then instead of ``throwing away'' the environment $\HS_f$,
apply a projector $F$ to this part of the output, and condition on the
resulting state.  One can think of this as carrying out a measurement on
$\HS_f$ that determines whether $F$ is true or false, and throwing away the
results of all experiments in which it is false. The consequence of an
appropriately chosen ``post selection'' of this type will be a set of
(conditional) probabilities that correspond to those of the original ket or
density operator; in other words, one obtains the same pre-probability --- see
Sec.~\ref{sct3} below.
The second approach is based on Fig.~\ref{fgr3}, and the idea is to
replace the fully-entangled $|\phi\rgl$ with a different entangled state,
chosen so $\phi_a$ is no longer proportional to $I_a$, but to 
$\Psi_a$ (or $R_a$). 

The question remains as to whether either of these procedures is worthwhile,
and that depends on one's goals.  Rather than turning entanglement problems
into channel problems, it may be simpler to do the reverse, as in the
classification scheme proposed in Sec.~\ref{sct6}.  This allows the
mathematical structure of the two types of problem to be compared.  If, on the
other hand, there is quite a bit of useful mathematical and physical intuition
to be wrung from contemplating how quantum systems develop in time, the
approaches mentioned in the previous paragraph may be worthwhile.  Until the
channel-entanglement duality has been more thoroughly explored, it is hard to
say which approach is best.  In any case, there are significant entanglement
problems that map in a simple way onto channel problems, and a study of what
entanglement does and does not mean in such cases might be very helpful.

\section{Quantum Information}
\label{sct3}

\subsection{Sample spaces and probabilities}
\label{sct3a}

The basic concept of ``information'' used in the following discussion is that
of a \emph{statistical correlation}.  This morning's newspaper contains
information because the symbols are correlated in an appropriate way with
yesterday's events.  Information is contained in a photon traveling down an
optical fiber because its properties are correlated with whatever produced it,
and with the effects produced by the further processes it will undergo.  An
encrypted message contains information in that its symbols are correlated with
those in the key used to encrypt or decrypt it.  Shannon's information theory
provides numerical measures for these statistical correlations, which apply to
quantum as well as to classical systems (which, of course, are in fact quantum
mechanical!), when probabilities have been properly defined.

Standard probability theory \cite{Fllr68,Rss00,DGSc02} is based on the idea of
a \emph{sample space} of mutually exclusive properties.  A quantum sample space
or \emph{framework} can be constructed using the mutually exclusive properties
associated with a decomposition of the identity (Sec.~\ref{sct2a}) of the
Hilbert space used to describe the system.  Given such a sample space one can
assign probabilities using the standard formula
\begin{equation}
  p_k = \lgl P^k\rgl = \lgl\psi|P^k|\psi\rgl = \Tr(P^k\rho),
\label{eqn31}
\end{equation}
where the quantum system is assumed to be described by a ket $|\psi\rgl$ or
density operator $\rho$ functioning as a \emph{pre-probability}, i.e., as a
device for generating probabilities \cite{Grff02b}.
Probabilities in quantum mechanics are often discussed in terms of
\emph{measurements}, which provide a good approach to understanding them in
operational terms, even though it is rather unsatisfactory from a fundamental
perspective (the infamous ``measurement problem''; see, e.g., \cite{Mttl98}).
For present purposes such measurements should be thought of as \emph{ideal
projective measurements} which reveal the (microscopic) properties they are
designed to measure; see the discussion in Chs.~17 and 18 of \cite{Grff02b}. We
shall have no need of more complicated concepts such as POVMs (see, e.g., p.~90
of \cite{NlCh00}, or Ch.~7 of \cite{dMyn02}).  From time to time there have
been proposals to introduce nonstandard notions of probability into quantum
mechanics, but these have not proven very successful, and we shall not use
them.

In quantum mechanics, in contrast to classical physics, one is typically
interested in a variety of sample spaces that are incompatible with each other,
but whose probabilities can all be generated from a single pre-probability.
For example, what is the probability that $S_x=+1/2$, or that $S_z=-1/2$, for a
spin-half particle?  The same ket or density operator may be used to answer
these questions by inserting different projectors in \eqref{eqn31}, but there
is no way of combining the answers to make them refer to a single physical
system, as it makes no sense to talk about $S_x=+1/2$ AND $S_z=-1/2$, or any
other logical combination of propositions associated with incompatible
decompositions of the identity whose projectors do not commute with each other.
Traditional textbooks state that $S_x$ and $S_z$ cannot be simultaneously
\emph{measured}, which is correct.  But the reason such joint measurements are
impossible in a quantum world is that the combined properties \emph{do not
exist}: such a combination is incompatible with the mathematical structure of
the quantum Hilbert space, see Ch.~4 of \cite{Grff02b}.  Treating incompatible
sample spaces as if they were compatible and combining the probabilities of one
with the other is the same sort of mistake as ignoring the difference between
$xp$ and $px$ when these symbols refer to quantum operators.

Consequently, one must be careful when giving a \emph{physical} interpretation
to the various \emph{mathematical} constraints, such as those in
Sec.~\ref{sct4}, relating probabilities on different incompatible sample spaces
generated by a single pre-probability.  They cannot refer to a single quantum
system, as it cannot be simultaneously described by incompatible frameworks.
Instead, one must take a counterfactual approach: ``This is what happens when a
qubit initially in state $|0\rgl$ is sent through the channel, \emph{but if
instead it had been} in the state $(|0\rgl+|1\rgl)/2$, then\dots.'' To be sure,
counterfactuals can themselves produce headaches in quantum theory if
improperly used; for a consistent approach, see Ch.~19 of \cite{Grff02b}.
Alternatively, one can imagine different experiments carried out on an array of
nominally identical systems.

 In comparison with classical physics, the new and unfamiliar element in
quantum information theory is the multiplicity of incompatible sample spaces
and probability distributions associated with them, even when one is using a
single pre-probability.  Finding good ways to think about this is a fundamental
problem, perhaps the fundamental problem, of quantum information, and thus a
major challenge to our understanding the world in quantum terms.

\subsection{Correlations}
\label{sct3b}

Consider two systems $\ST_a$ and $\ST_b$, with Hilbert spaces
$\HS_a$ and $\HS_b$, and let $\{A^j\}$ and $\{B^k\}$ be decompositions
of the respective identities $I_a$ and $I_b$.  On the tensor product 
$\HS_{ab}=\HS_a\ot\HS_b$ used to describe the combined systems the projectors
$\{A^j B^k\}$ form a decomposition of $I_{ab}$, and thus a sample space, to
which probabilities may be assigned as in \eqref{eqn31}:
\begin{equation}
  \Pr(A^j,B^k) = \lgl A^j B^k\rgl = \Tr\left[(A^j\ot B^k)\rho\right],
\label{eqn32}
\end{equation}
with $\rho = |\psi\rgl\lgl\psi|$ for a pure state $|\psi\rgl$.  The marginal
distributions
\begin{gather}
  \Pr(A^j)  = \sum_k \Pr(A^j,B^k) = \lgl A^j\rgl,
\notag\\
  \Pr(B^k)  = \sum_j \Pr(A^j,B^k) = \lgl B^k\rgl
\label{eqn33}
\end{gather}
are obtained by summing or by inserting $A^j\ot I$ (i.e., $A^j$) or $I\ot B^k$
(i.e., $B^k$) on the right side of \eqref{eqn32}. One can think of
$\Pr(A^j,B^k)$ as the joint probability distribution of two random variables
which take on integer values $j$ and $k$, and apply to it any standard measure
of correlation including, if one wants, the Shannon mutual information
$I(A\colo B)$. Note, in particular, the condition for \emph{statistical
independence}:
\begin{equation}
  \Pr(A^j,B^k) =\Pr(A^j)\Pr(B^k), \text{ or } 
    \lgl A^j B^k\rgl = \lgl A^j\rgl\lgl B^k\rgl.
\label{eqn34}
\end{equation}

If one thinks of $\ST_a$ and $\ST_b$ as physically separated systems, then the
joint probability distribution \eqref{eqn32} will be the same as that of the
outcomes of ideal measurements of $\{A^j\}$ and $\{B^k\}$ carried out on the
separate systems. Consequently, the measurement outcomes will be correlated in
precisely the same way as the quantum properties that have been measured, and
one can use either the language of properties (our approach) or of
measurement outcomes to discuss these statistical correlations.
Discussions of measurements in textbooks often refer to ``observables'' rather
than decompositions.  Given a decomposition $\{A^j\}$, one can always construct
a corresponding observable $O=\sum_j a_j A^j$ with distinct (real) eigenvalues:
$a_j\neq a_k$ for $j\neq k$ .  But for our purposes these eigenvalues play no
role, so the language of decompositions tends to be clearer than that referring
to observables.

\subsection{Information present and absent}
\label{sct3c}

Because of the multiplicity of incompatible quantum sample spaces, one needs to
identify different \emph{types} or \emph{varieties} of information potentially
available about a particular system.  Given a decomposition $\AS=\{A^j\}$ of
$I_a$, we shall say that the $\AS$ information about $\ST_a$ is \emph{present},
or \emph{perfectly present}, in another system $\ST_b$ for a given
pre-probability provided there exists a decomposition $\BS=\{B^k\}$ of $I_b$
such that
\begin{equation}
  \lgl A^j B^k\rgl = \dl_{jk}\lgl A^j\rgl = \dl_{jk}\lgl B^k\rgl,
\label{eqn35}
\end{equation}
where one may have to renumber the projectors in one of the collections to
satisfy this condition.  A little thought will show that the first equality
implies the second.  The symmetry of the definition implies that when some type
of information about $\ST_a$ is available in $\ST_b$, there is also some type
of information about $\ST_b$ available in $\ST_a$. Although we shall not make
use of it in this paper, it is worth mentioning that the Shannon mutual
information $I(\AS\colo \BS)$ in this case is $(-\sum_j p_j\log p_j)$ with $p_j
= \lgl A^j\rgl$.

If the $\AS$ information about $\ST_a$ is present in $\ST_b$ (in the sense just
defined) for \emph{every} decomposition of $I_a$, we shall say that \emph{all}
the (quantum) information about $\ST_a$ is in $\ST_b$.  Clearly it suffices to
check this for every orthonormal basis $\{|a^j\rgl\}$.  Less obvious
(theorem~\ref{thm4} in Sec.~\ref{sct4}) is the fact that one need not
check them all: two properly chosen incompatible bases  suffice.  We shall
say that $\ST_a$ and $\ST_b$ are \emph{informationally equivalent} when all
information about $\ST_a$ is in $\ST_b$ and all information about $\ST_b$ is in
$\ST_a$.

The $\AS=\{A^j\}$ information about $\ST_a$ is (completely) \emph{absent} from 
$\ST_b$ provided \emph{any} choice of a decomposition $\{B^k\}$ of $I_b$ is
statistically independent, \eqref{eqn34}.  A little thought shows that this
is equivalent to the requirement that
\begin{equation}
  \Tr_a(A^j\rho) = \lgl A^j\rgl\rho_b = p_j\rho_b
\label{eqn36}
\end{equation}
for every $j$, where $\rho_b=\Tr_a(\rho)$ is the reduced density operator for
$\rho_b$.  (Note that it suffices to require that the operators defined by the
left side of \eqref{eqn36} be proportional to one another; when that is so,
summing them shows they are all proportional to $\rho_b$.)  In other words, for
every $j$ such that $p_j$ is not zero, the density operator \emph{conditional}
on $A^j$,
\begin{equation}
  \bar\rho_b^j = \Tr_a(A^j\rho)/p_j,
\label{eqn37}
\end{equation}
is the same as $\rho_b$.

If for every decomposition $\AS$ of $I_a$ --- it suffices to check all
orthonormal bases --- the corresponding information about $\ST_a$ is absent
from $\ST_b$, one can show (theorem~\ref{thm1} (iii) in Sec.~\ref{sct4}) that
\begin{equation}
  \rho = \rho_a\ot\rho_b,
\label{eqn38}
\end{equation}
from which it follows that all information of any sort about $\ST_b$ is also
absent from $\ST_a$.  In this case we shall say that $\ST_a$ and $\ST_b$ are
(completely) \emph{uncorrelated}.  No conceivable measurement on one of these
systems will provide any information about the other. 

In the case of three or more systems, the presence or absence of particular
types of information about $\ST_a$ satisfies some intuitively obvious rules. If
$\AS$ information about $\ST_a$ is present in $\ST_b$, it is also present in
the combined system $\ST_b$ and $\ST_c$, denoted by $\ST_{bc}$. If it is absent
from $\ST_{bc}$, it is absent from both $\ST_b$ and $\ST_c$.  The same is true
when ``$\AS$ information'' is replaced by ``all information.''

These definitions of information perfectly present or completely absent make no
reference to any sort of numerical measure of correlation, and thus are useful
for a \emph{qualitative} rather than a quantitative discussion of quantum
information.  This is not to say that quantitative measures are unimportant ---
far from it --- but they lie outside the scope of this paper. It is hoped that
the qualitative approach developed here will help organize and motivate
quantitative discussions, see Sec.~\ref{sct7b}.

\subsection{Correlations for channels}
\label{sct3d}

The preceding discussion referred to properties of separated systems $\ST_a$
and $\ST_b$ at the same time.  Basically the same ideas apply in the
case of quantum channels, where $\ST_a$ is the channel input at an earlier time
and $\ST_b$ its output at a later time (Sec.~\ref{sct2c}).  The only difference
is the manner in which one calculates a joint probability distribution;
\eqref{eqn32} is replaced by
\begin{equation}
   \Pr(A^j,B^k) = \lgl A^j B^k\rgl = \Tr\left[(A^j\ot B^k)Q\right].
\label{eqn39}
\end{equation}
Here the transition operator $Q$, see \eqref{eqn23}. takes the place of the
density operator in \eqref{eqn32}.  The marginals are once again given by
\eqref{eqn33}.  The fact that $Q$ is the partial transpose of a density
operator $R$ guarantees that the probabilities in \eqref{eqn39} are well
defined; indeed, they behave very much like those of a bipartite system
described by $R$.  

One can once again visualize $\{B^k\}$ in terms of idealized measurements of
what emerges from the channel, but the corresponding intuitive picture of
$\{A^j\}$ is an ideal \emph{preparation}.  Of course, it is no more possible to
prepare a quantum system in a state of two (or more) incompatible properties
than it is to measure such a state, for such states do not exist in the quantum
world.  And just as an ideal measurement reveals a property possessed by a
quantum system at a slightly earlier time, an ideal preparation results in a
quantum system having a specific property at a slightly later time.
The language of ``preparation'' and ``measurement'' is useful both for
providing quantum concepts with intuitive content and for relating quantum
theory to laboratory experiments, but it should be used to illuminate, not
replace, the notion of statistical correlations among microscopic properties,
whether at the same or at different times, as this is the more fundamental
concept.

The correlations obtained using a transition operator, \eqref{eqn39}, are not
entirely the same as those arising from a density operator, \eqref{eqn32}, but
the differences are rather subtle.  Given a pair of decompositions $\AS$ and
$\BS$, there is no way of telling whether the joint probability distribution
comes from a density or a transition operator.  What can happen with
\emph{sets} of correlations for \emph{incompatible} decompositions, when they
are generated by a single pre-probability, is best illustrated by means of an
example.  For a perfect one-qubit channel, $p=0$ in \eqref{eqn26}, each
component of angular momentum of a spin-half particle is identical at the
entrance and at the exit,
\begin{equation}
 \lgl \sg_a^x\sg_b^x\rgl = \lgl\sg_a^y\sg_b^y\rgl = \lgl\sg_a^z\sg_b^z\rgl = 1.
\label{eqn40}
\end{equation}
However, this type of correlation is impossible for two separate systems at the
same time. What one can, instead, achieve by using an appropriate (pure state)
density operator is
\begin{equation}
 \lgl \sg_a^x\sg_b^x\rgl= -\lgl\sg_a^y\sg_b^y\rgl = \lgl\sg_a^z\sg_b^z\rgl = 1
\label{eqn41}
\end{equation}
or something similar: one of the terms (it need not be
$\lgl\sg_a^y\sg_b^y\rgl$) must have a minus sign, or else there are three minus
signs, as in the famous spin-singlet state used in discussions of the
Einstein-Podolsky-Rosen paradox. Similarly, \eqref{eqn41} is impossible for a
quantum channel.

Interesting as these differences, which arise from the partial transpose in
\eqref{eqn23}, may be, they are basically irrelevant to the concerns of this
paper.  The definitions of information perfectly present or completely absent
given in Sec.~\ref{sct3c} above and the theorems in Sec.~\ref{sct4} below apply
equally to channels and entangled states.  In both cases the fundamental issue
is statistical correlations and what quantum theory has to say about them, and
that is exactly the same once proper account is taken of the partial transpose.

\section{All or Nothing Theorems}
\label{sct4}

It is convenient to organize a number of qualitative ``all or nothing'' results
on the location of quantum information in a series of eight theorems.  The
first four refer to bipartite and the last four to tripartite systems.  In
several cases there are separate results depending upon whether the
pre-probability is a pure state, indicated by a ket $|\Psi\rgl$, or a density
operator $\rho$. The former are stronger than the latter, and the reader should
keep in mind that any result that is valid for a density operator applies
equally to the case of a pure state, even if that is not explicitly stated.

While the theorems are stated for entangled states, thought of as different
systems at a single instant of time, they apply equally to correlations at two
different times in a quantum channel, for which $|\Psi\rgl$ is the channel
ket.
The bipartite systems used in the first four theorems are sometimes designated
$\ST_{ab}$ and sometimes $\ST_{ac}$.  This makes the notation consistent with
the later theorems for tripartite systems, where information \emph{about}
$\ST_a$ is \emph{present} in $\ST_b$ and/or \emph{absent} from $\ST_c$.  Note
that, in agreement with the definitions in Sec.~\ref{sct3c}, ``present'' means
perfectly or completely present; ``absent'' means completely absent.  The
proofs will be found in App.~A.

The tripartite theorems have a no-cloning ``smell'' to them, and represent an
attempt to give this important, but somewhat elusive, notion a precise
information-theoretic content.  The absence of theorems for $p$-part systems
with $p\geq 4$ reflects our inability to find results of corresponding
generality, and we hope our readers will be more successful.  But keep in mind
that a tripartite theorem might, for example, be usefully applied to
$\ST_{abcd}$ thought of as consisting of $\ST_a$, $\ST_b$, and $\ST_{cd}$ --- a
strategy employed in discussing quantum codes in Sec.~\ref{sct5c}.

\begin{theorem} 

Absence of information.

i) If $\AS=\{A^l\}$ is a decomposition of $I_a$, the $\AS$ information about
$\ST_a$ is absent from $\ST_c$ for a pre-probability $|\Psi\rgl\,\in \HS_{ac}$
if and only if
\begin{equation}
  P A^l P = a_l P,
\label{eqn42}
\end{equation}
where $P$ is the projector on the support of $\Psi_a$, and the $a_l$ are
(nonnegative) constants.  
The following is equivalent to \eqref{eqn42}:
\begin{equation}
  \lgl p^j |A^l|p^k\rgl = a_l\dl_{jk},
\label{eqn43}
\end{equation}
where $\{|p^j\rgl\}$ is a collection of orthonormal states
which span the support of $\Psi_a$, so that $P=\sum_j |p^j\rgl\lgl p^j|$.

ii) If $\AS=\{|a^j\rgl\}$ is an orthonormal basis and all $\AS$ information 
about $\ST_a$ is absent from $\ST_c$ for $|\Psi\rgl\,\in \HS_{ac}$, then
\begin{equation}
  |\Psi\rgl = |\al\rgl\ot|\gm\rgl
\label{eqn44}
\end{equation}
is a product state on $\HS_a\ot\HS_c$. 

iii) All information about $\ST_a$ is absent from $\ST_c$ for a pre-probability
$\rho\,\in\hat\HS_{ac}$ if and only if 
\begin{equation}
  \rho= \rho_a\ot\rho_c,
\label{eqn45}
\end{equation}
which implies that all information about $\ST_c$ is absent from $\ST_a$ (the
two are uncorrelated). 
 
\label{thm1}
\end{theorem}

\begin{theorem}

Presence of particular information.
%

i) The $\AS=\{A^l\}$ information about  $\ST_a$ is present in $\ST_b$ for
$\rho\in\hat\HS_{ab}$ if and only if
\begin{equation}
  \Lm^l \Lm^m=0 \text{ for } l\neq m,
\label{eqn46}
\end{equation}
where
\begin{equation}
 \Lm^l = \Tr_a(A^l\rho). 
\label{eqn47}
\end{equation}

ii) The $\AS=\{A^l\}$ information about $\ST_a$ is present in $\ST_b$ for
$|\Psi\rgl\,\in\HS_{ab}$ if and only if
\begin{equation}
  [A^l,\Psi_a]=0
\label{eqn48}
\end{equation}
for all $l$.  In particular, if $\AS=\{|a^j\rgl\}$ is an orthonormal basis,
\eqref{eqn48} is equivalent to the requirement that
\begin{equation}
  |\Psi\rgl = \sum_j |a^j\rgl\ot|\bt^j\rgl
\label{eqn49}
\end{equation}
be a Schmidt expansion, i.e., $\lgl\bt^k|\bt^j\rgl=0$ for $j\neq k$. 

iii) If the $\AS=\{A^l\}$ information about $\ST_a$ is present in $\ST_b$ for
$\rho\,\in\hat\HS_{ab}$, then for all $l$
\begin{equation}
  [A^l,\rho_a]=0.
\label{eqn50}
\end{equation}
\label{thm2}
\end{theorem}

Note that if $\AS$ is an orthonormal basis, \eqref{eqn48} and \eqref{eqn50}
are equivalent to the assertion that the $\Psi_a$ or $\rho_a$ matrices are
diagonal in this basis.

\begin{theorem}

Presence of all information. 

i) All information about $\ST_a$ is in $\ST_b$ for $|\Psi\rgl\,\in\HS_{ab}$
if and only if
\begin{equation}
  \Psi_a=I_a/d_a,
\label{eqn51}
\end{equation}
i.e., $|\Psi\rgl$ is maximally entangled.

ii) All information about $\ST_a$ is in $\ST_b$ for $\rho\,\in\hat\HS_{ab}$
if and only if there are Hilbert spaces $\HS_d$ and $\HS_e$ whose tensor
product is $\HS_b$ or a subspace of $\HS_b$, and $\rho$ is of the form
\begin{equation}
  \rho = \phi\ot\rho_e \,\in\hat\HS_{ad}\ot\hat\HS_e,
\label{eqn52}
\end{equation}
where $\phi = |\phi\rgl\lgl\phi|$ projects on a fully-entangled state
$|\phi\rgl\,\in\HS_{ad}$. This last implies (but is not implied by)
\begin{equation}
  \rho_a=I_a/d_a.
\label{eqn53}
\end{equation}

iii) All information about $\ST_a$ is in $\ST_b$ and all information
about $\ST_b$ is in $\ST_a$, i.e., the two systems are informationally
equivalent, if and only if the pre-probability is a fully-entangled pure state,
i.e., maximally entangled with $\HS_a$ and $\HS_b$ of the same dimension.
\label{thm3}
\end{theorem}

The utility of theorem~\ref{thm3} increases significantly through the existence
of some (seemingly) rather weak conditions which imply that all information
about $\ST_a$ is in $\ST_b$.  To this end we need the following definition.
Two decompositions $\AS=\{A^j\}$ and $\bar\AS=\{\bar A^k\}$ of $I_a$ are
\emph{strongly incompatible} if there exists no projector $P$, apart from $P=0$
and $P=I_a$, that commutes with all the $\{A^j\}$ and all the $\{\bar A^k\}$.
This is, for example, the case when $\AS=\{|a^j\rgl\}$ and $\bar\AS=\{|\bar
a^j\rgl\}$ are two orthonormal bases for which
\begin{equation}
  \lgl a^j|\bar a^k\rgl \neq 0
\label{eqn54}
\end{equation}
for all $j$ and $k$, a condition which is fulfilled when the two bases are
mutually unbiased, \eqref{eqn4}, but is obviously much weaker.  Strong
incompatibility is weaker still; it is possible for a number of the inner
products in \eqref{eqn54} to vanish provided a sufficient number are nonzero.
Indeed, two decompositions can be strongly incompatible without all of the
projectors, or, in some cases, any of the projectors being onto pure states. We
shall not pursue the matter further at this point, but instead state the
desired result:

\begin{theorem}

Strong incompatibility. Let $\AS$ and $\bar\AS$ be two strongly incompatible
decompositions of $I_a$, according to the preceding definition, and suppose
that both the  $\AS$ and the $\bar\AS$ information about $\ST_a$ is in
$\ST_b$.  Then 
\begin{equation}
  \rho_a = I_a/d_a,
\label{eqn55}
\end{equation}
and if, in addition, $\rho=\Psi$ is a pure state on $\HS_{ab}$, then
\emph{all} information about $\ST_a$ is in $\ST_b$. 
\label{thm4}
\end{theorem}

The following theorems refer to a tripartite system $\ST_{abc}$.

\begin{theorem}

All information absent.  If for $|\Psi\rgl\,\in\HS_{abc}$ all information about
$\ST_a$ is absent from $\ST_c$, there are Hilbert spaces $\HS_d$ and $\HS_e$
whose tensor product $\HS_{de}$ is either $\HS_b$ or a subspace of $\HS_b$,
and $|\Psi\rgl$ is of the form
\begin{equation}
  |\Psi\rgl = |\chi\rgl\ot|\psi\rgl \,\in \HS_{ad}\ot\HS_{ce}.
\label{eqn56}
\end{equation}
\label{thm5}
\end{theorem}

Only if the support of $\Psi_b$ is a proper subspace of $\HS_b$ will $\HS_{de}$
differ from $\HS_b$, and in that case it can be identified with the subspace.
The ``hidden product'' structure of \eqref{eqn56} turns out to be a
surprisingly useful tool.

\begin{theorem}

Particular information present for a pure state.
For a pre-probability $|\Psi\rgl\in\HS_{abc}$:

i) If $\AS=\{|a^j\rgl\}$ is an orthonormal basis of $\HS_a$,
a necessary and sufficient condition for the $\AS$ information
to be present in $\ST_b$ is that
\begin{equation}
  \Psi_{ac} = \sum_j |a^j\rgl\lgl a^j|\ot \Gm^j,
\label{eqn57}
\end{equation}
where the $\{\Gm^j\}$ are (positive) operators on $\HS_c$.

\vspace{.3ex}

ii) If for some decomposition $\AS=\{A^k\}$ of $I_a$,
\begin{equation}
  \Psi_{ac} = \sum_k A^k\ot\bar\Gm^k,
\label{eqn58}
\end{equation}
the $\AS$ information about $\ST_a$ is in $\ST_b$, and if $\bar\AS=\{\bar
A^l\}$ is a compatible decomposition of $I_a$ in the sense that all the $\{\bar
A^l\}$ projectors commute with all the $\{A^k\}$ projectors, then the $\bar\AS$
information is also present in $\ST_b$.  (In particular, $\bar\AS$ may be an
orthonormal basis in which the $\{A^k\}$ are diagonal.)
\label{thm6}
\end{theorem}

\begin{theorem}

Particular information present for a mixed state.
Suppose that the $\AS=\{|a^j\rgl\}$ information about $\ST_a$ is in $\ST_b$
for $\rho\in\hat\HS_{abc}$. Then

i) The reduced density operator on $\HS_{ac}$ is of the form
\begin{equation}
  \rho_{ac} = \sum_j |a^j\rgl\lgl a^j|\ot \Gm^j,
\label{eqn59}
\end{equation}
where the $\{\Gm^j\}$ are (positive) operators on $\HS_c$.

ii) If $\bar\AS=\{|\bar a^k\rgl\}$ is another orthonormal basis of $\HS_a$, and
$\AS$ and $\bar \AS$ are mutually
unbiased, then no $\bar\AS$ information is in $\ST_c$, and
\begin{equation}
  \Tr(\rho[\bar a^k]) = 1/d_a,
\label{eqn60}
\end{equation}
independent of $k$. 
\label{thm7}
\end{theorem}

\begin{theorem}

No splitting theorem.

i) If for $\rho\,\in\hat\HS_{abc}$ all the information about $\ST_a$ is in
$\ST_b$, then there is no information about $\ST_a$ in $\ST_c$,
\begin{equation}
  \rho_{ac}=\rho_a\ot\rho_c.
\label{eqn61}
\end{equation}

ii) If for $|\Psi\rgl\,\in\HS_{abc}$ all the information about $\ST_a$ is in
$\ST_{bc}$, and none of it is in $\ST_c$, then it is all in $\ST_b$. 

iii) If for $\rho\in\hat\HS_{abc}$ all the information about $\ST_a$ is in
$\ST_{bc}$, but none of it is in $\ST_c$, then the dimension of $\HS_b$ is not
less than that of $\HS_a$. 
\label{thm8}
\end{theorem}

Note that (iii) in this last theorem is a weaker result than (ii), for if all
the $\ST_a$ information is in $\ST_b$, then by theorem~\ref{thm3} (ii) the
dimension of $\HS_b$ cannot be less than that of $\HS_a$.  The difference
between (ii) and (iii) turns out be of some interest for understanding
quantum codes, Sec.~\ref{sct5c}.

\section{Applications}
\label{sct5}

\subsection{Channels with mixed-state environment}
\label{sct5a}

There is no loss in generality in assuming the environment for a quantum
channel is initially in a pure state, Fig.~\ref{fgr2}, provided the dimension
$d_e$ of $\HS_e$ is at least $d_a^2$.  The question has been raised
\cite{Trao99,ZlRf02} as to what channels can be produced using a smaller $d_e$,
e.g., $d_e=d_a$, if one assumes an initial mixed state for the environment.
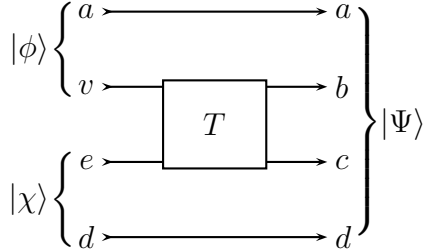
\begin{figure}[h]
$$
\begin{pspicture}(-1.0,-1.2)(4.0,2.2) 
\def\rectc(#1,#2){%
\psframe[fillcolor=white,fillstyle=solid](-#1,-#2)(#1,#2)}
\def\vertdash(#1){\psline[linestyle=dashed,linewidth=0.01](0.0,0.0)(0.0,#1)}
\def\vtpair{\vertdash(1.0)\rput(0.1,0){\vertdash(1.0)}%
\psline(0.0,0.0)(0.1,0.0)\psline(0.0,1.0)(0.1,1.0)}
\def\vvv#1{\vrule height #1 cm depth #1 cm width 0pt}
\def\rbrac{$\left.\vvv{1.5}\right\}$}
\def\lbrac{$\left\{\vvv{0.5}\right.$}
\psline{>->}(0.0,2.0)(3.0,2.0)
\psline{>->}(0.0,1.0)(3.0,1.0)
\psline{>->}(0.0,00)(3.0,00)
\psline{>->}(0.0,-1.0)(3.0,-1.0)
\rput(1.5,0.5){\rectc(0.7,0.6)}
\rput(1.5,0.5){$T$}
\rput(-0.5,1.5){\lbrac}
\rput(-0.5,-0.5){\lbrac}
\rput[r](-0.1,2.0){$a$} 
\rput[r](-0.1,1.0){$v$} 
\rput[r](-0.1,0.0){$e$} 
\rput[r](-0.1,-1.0){$d$} 
\rput[r](-0.7,1.5){$|\phi\rgl$}
\rput[r](-0.7,-0.5){$|\chi\rgl$}
\rput[l](3.1,2.0){$a$}
\rput[l](3.1,1.0){$b$}
\rput[l](3.1,0.0){$c$}
\rput[l](3.1,-1.0){$d$}
\rput[l](3.3,0.5){\rbrac}
\rput[l](3.7,0.5){$|\Psi\rgl$}
\end{pspicture}
$$
\caption{%
Channel and channel ket $|\Psi\rgl$ for a mixed-state environment.}
\label{fgr4}
\end{figure}

Such a channel can be modeled in the manner indicated in Fig.~\ref{fgr4}, with
a ``large'' environment $\ST_{ed}$ initially in a pure state $|\chi\rgl$, which
when traced down to $\HS_e$ yields the desired mixed-state density operator.
The unitary transformation $T$ maps $\HS_{ve}$ onto $\HS_{bc}$ to produce the
analog of Fig.~\ref{fgr3}, where $f$ has become the pair $cd$, and $|\phi\rgl$
is again the fully-entangled state \eqref{eqn11}. The channel ket
\begin{equation}
  |\Psi\rgl = \left(I_a\ot T\ot I_d\right) (|\phi\rgl\ot|\chi\rgl
\label{eqn62}
\end{equation}
is a pure state of $\HS_{abcd}$.

This channel ket has the interesting property
\begin{equation}
  \Psi_{ad} = \Psi_a\ot\Psi_d,
\label{eqn63}
\end{equation}
which means that $\ST_a$ and $\ST_d$ are uncorrelated; no information about one
is available in the other.  It follows from the fact that the product state on
the right side of \eqref{eqn62} has this property, which is preserved during
time development because the \emph{unitary} operator $T$ does not act on
$\HS_{ad}$.  As a consequence, $\Psi_{ad}$ (and therefore also its partial
traces $\Psi_a$ and $\Psi_d$) is independent of time.  Note that this
invariance is \emph{not} true (in general) if $T$ is not a unitary operator.
The reason, in physical terms, is that a general map from $\HS_{ve}$ to
$\HS_{bc}$ can be thought of as involving post selection, based upon some sort
of joint measurement.  Since $\ST_a$ is correlated with $\ST_z$ and $\ST_d$
with $\ST_e$ through the entangled initial states, the final state of affairs
\emph{conditioned on the outcome of such a measurement} may very well contain
correlations between $\ST_a$ and $\ST_d$.

Not only is \eqref{eqn63} a consequence of our model of a mixed-state
environment, it comes close to being the very essence of the matter in light of
theorem~\ref{thm5} applied to the tripartite $\HS_a\ot\HS_{bc}\ot\HS_d$, for
that tells us that $|\Psi\rgl$ necessarily involves a ``hidden product''
structure.  What is required to bring that structure to light is a suitable
unitary transformation, which is $T$ in Fig.~\ref{fgr4}.  To be sure,
theorem~\ref{thm5} does not tell us that $|\phi\rgl$ shall be fully entangled
--- which suggests that the problem of a channel with a mixed-state environment
is actually part of a more general information-theoretical question about
entangled states on 4-part systems, and exploring it from this perspective may
be useful.  In addition, our analysis suggests a close connection between such
channels and properties of unitary transformations on bipartite systems.  

\subsection{ CQ  channels}
\label{sct5b}

The notion of a  CQ  or ``classical-quantum'' channel was introduced in
\cite{Hlvo99}, and has been the subject of some recent studies
\cite{HrSR03,Rska03} in connection with entanglement-breaking channels, which
were introduced in \cite{Shr02}.  An entanglement-breaking channel may be
defined as one in which the dynamical operator $R$ in \eqref{eqn21} is
separable, in the standard way in which that term is applied to density
operators (see, e.g., \cite{Lwao00,HrHH01}, Sec.~2.2.3 of \cite{Kyl02}),
and a  CQ  channel is a particular case of an entanglement-breaking channel
in which $R$ has the form
\begin{equation}
  R = (1/d_a)\sum_j |a^j\rgl\lgl a^j|\ot B^j,
\label{eqn64}
\end{equation}
using a suitably chosen orthonormal basis $\AS=\{|a^j\rgl\}$ for $\HS_a$, and
positive operators $B^j$ of unit trace (to ensure \eqref{eqn15}) on $\HS_b$.
The remarks which follow apply equally to a QC or ``quantum-classical''
channel, with the roles of $a$ and $b$ interchanged.

Introducing the channel ket $|\Psi\rgl\,\in\HS_{abf}$ with $R=\Psi_{ab}$,
\eqref{eqn14}, allows one to apply theorem~\ref{thm6} (i) in order to
characterize a  CQ  channel as one in which there is an orthonormal basis for
the channel entrance such that the information associated with this basis is
\emph{perfectly present in the environment} $\ST_f$ at the later time.  Note
that such a characterization is \emph{not} immediately obvious from considering
the dynamical operator, or, equivalently, the channel superoperator, for these
are obtained by tracing out, thus ignoring, the environment, whereas the
property which provides the simplest characterization in information-theoretic
terms has very much to do with what information is available \emph{in} the
environment!

Using a channel ket in no way reduces the value of the insights provided in the
studies cited above, nor does it supply (at least in any obvious sense)
alternative tools for arriving at the technical results in those papers.  But
it does suggest a genuinely quantum-mechanical and information-theoretical
description of what is ``classical'' (the C in  CQ ) about a  CQ  channel:
namely, the environment provides perfect decoherence in a particular basis, as
a consequence of which no information in any ``complementary'', which is to say
mutually unbiased basis, is available at the channel exit, theorem~\ref{thm7}
(ii).  This is typical of what is generally referred to as ``classical
communication.''

\subsection{Information location in quantum codes}
\label{sct5c}

Quantum codes allow quantum information to be preserved against the effects of
noise, whether due to interaction with the environment in a quantum
communication setting, or imperfect gates in a quantum computer, and thus they
have received a great deal of attention; for an introduction, see \cite{Stne98}
and Ch.~10 of \cite{NlCh00}.  Our purpose here is not to contribute to the
technical literature, but instead to point out how the basic operation of such
a code can be understood in terms of the presence or absence of certain types
of information in certain places.

The standard scenario is one in which the quantum information is embedded in
a  \emph{code} $\BS$, a $K$-dimensional subspace of the Hilbert space
\begin{equation}
  \HS_d = \HS_1\ot\HS_2\ot\cdots \HS_n
\label{eqn65}
\end{equation}
associated with $n$ \emph{carriers} of the coded information. The simplest
situation is one in which $K=2=d_m$ for $1\leq m\leq n$, but most of what we
have to say applies more generally.
Define the \emph{security} $s$ of the code to be the largest integer such that
the encoded information is entirely absent from any set of $s$ or fewer
carriers (in a sense made precise in \eqref{eqn68} below).  That is, an
eavesdropper could learn nothing at all by carrying out arbitrary measurements
on a set of $s$ carriers, but could learn something from a suitable set of
$s+1$ carriers.  In the literature it is customary to refer to $s+1$ as the
``distance'' $d$ of the code, using an analogy with classical codes in which
$d$ is the minimum Hamming distance between two code words.  For a quantum code
the notion of ``distance'' is somewhat obscure, as is the notion of code word,
whereas $s$ has a simple intuitive interpretation.

For analyzing the security and the error-correction properties of the
code it is convenient to define a channel ket
\begin{equation}
  \sqrt{K}\,|\Psi\rgl =\sum_j |a^j\rgl \ot |b^j\rgl\,\in\HS_a\ot\HS_d, 
\label{eqn66}
\end{equation}
where the $\{|a^j\rgl\}$ form an orthonormal basis of the channel entrance
$\HS_a$, with $d_a=K$, and the $\{|b^j\rgl\}$ an orthonormal basis of the 
code subspace $\BS$ with projector 
\begin{equation}
  B=\sum_j |b^j\rgl\lgl b^j|.
\label{eqn67}
\end{equation}
Thus the encoding operation maps $\HS_a$ onto $\BS$.  One can visualize
$|\Psi\rgl$ using Fig.~\ref{fgr3}, but with $\ST_b$ and $\ST_f$ combined to
form $\ST_d$.

The security condition introduced earlier can now be stated as
\begin{equation}
  \Psi_{au} = \Psi_a\ot\Psi_u,
\label{eqn68}
\end{equation}
where $u$ denotes any subset of $s$ integers drawn from $\{1,2,\ldots n\}$.
Note that if \eqref{eqn68} holds for such a set, it also holds for a smaller
set; simply take an appropriate partial trace of both sides.  In view of
theorem~\ref{thm1} (iii), \eqref{eqn68} expresses precisely what we want to say
by the security condition: if it is satisfied, no conceivable measurement on
$\ST_u$ will reveal anything about any sort of information in the channel
entrance, whereas if it is \emph{not} satisfied, \emph{some} sort of
information will be at least partially available to an eavesdropper.

From the definition \eqref{eqn66} it is obvious that $\Psi_a=I_a/d_a$, so by
theorem~\ref{thm3} (i) all information about $\ST_a$ is in $\ST_d$.  Thus by
theorem~\ref{thm8} (ii), if none of this information is in $\ST_u$, it must be
in the complement of this system in $\ST_d$.  That is, all the information
about $\ST_a$ is available in any collection of $n-s$ carriers; given any such
a set, there will be a means of extracting or recovering the information from
it even if the other carriers are ignored.  This provides a preliminary
understanding in information-theoretic terms of how a quantum error-correcting
code functions, though some additional points remain to be dealt with.

In order to relate the security of the code to the discussion of error
correction found in \cite{Bnao96,KnLf97}, it is helpful to introduce the
following definition.  An operator $F$ on $\HS_d$ will be said to have a
\emph{base} $\ST_w$, where $w$ is some subset of, and $\tilde w$ its complement
in, $\{1,2,\ldots n\}$, provided
\begin{equation}
  F = F_w\ot I_{\tilde w}\,\in \hat\HS_w\ot\hat\HS_{\tilde w},
\label{eqn69}
\end{equation}
and $w$ is the \emph{smallest} set for which $F$ can be
written in this form.  The \emph{size} of the base of $F$ is the number of
carriers in $\ST_w$, the number of integers in $w$.  


A code has security $s$ when for every operator $F$ with
a base whose size does not exceed $s$ it is the case that
\begin{equation}
  B F B =  b(F) B, \text{ or } \lgl b^j |F|b^k\rgl =  b(F)\dl_{jk},
\label{eqn70}
\end{equation}
and $s$ is the largest integer for which this is the case.  Here $b(F)$ is a
(complex) number that depends upon $F$, but not on $j$ or $k$, and $B$ is the
projector in \eqref{eqn67}.  The two equalities in \eqref{eqn70} are equivalent
because the second is simply the first expressed as a matrix when one extends
$\{|b^j\rgl\}$ to an orthonormal basis of $\HS_d$.  To see that \eqref{eqn70}
is correct, first apply it in the case where $F$ is a projector in $\hat\HS_u$
for some $u$ for which \eqref{eqn68} holds, and use theorem~\ref{thm1} (i),
with $\HS_a$ in the theorem replaced by $\HS_d$, and $\HS_c$ by $\HS_a$, to the
decomposition $\{F,I_d-F\}$ of $I_d$.  Any operator on $\HS_u$ can be expressed
as a linear combination of projectors, and hence by linearity, and the ``if and
only if'' of theorem~\ref{thm1} (i), we arrive at the equivalence of
\eqref{eqn68} and \eqref{eqn70} as statements that $\HS_a$ and $\HS_u$ are
uncorrelated.

Now \eqref{eqn70} is very similar to the necessary and sufficient condition
\begin{equation}
  \lgl b^j |K\ad_l K_m|b^k\rgl = b_{lm}\dl_{jk}
\label{eqn71}
\end{equation}
of \cite{KnLf97} (in a slightly different notation) for a code to be able to
correct a class of errors corresponding to the Kraus operators $\{K_l\}$ acting
on the space $\HS_d$.  If these Kraus operators have a base no larger than $t$,
then $F=K\ad_l K_m$ has a base that is no larger than $2t$, and we arrive at
the condition
\begin{equation}
  s=2t
\label{eqn72}
\end{equation}
relating the security $s$ to the maximum number of errors $t$ which can be
corrected.  That is, a code which allows full recovery of information when $t$
carriers are tampered with in any way, and \emph{one does not know which}
carriers have been affected, must allow full recovery when any \emph{known} set
of $s=2t$ carriers have been tampered with; in the latter case the information
will be recovered from the $n-2t$ remaining carriers.  Thus the well-known five
qubit code --- see \cite{Bnao96,LMPZ96} and p.~469 of \cite{NlCh00} --- allows
error recovery in the case of tampering with any one of the five carriers, but
also if any two are stolen, since recovery is then carried out on the three
that remain. (For a helpful discussion of this somewhat confusing point, 
see \cite{GrBP97}.) 

The foregoing considerations make it possible to understand in
information-theoretical terms the quantum Singleton lower bound
\begin{equation}
  n\geq 4t + \log K/\log D
\label{eqn73}
\end{equation}
on the number of carriers, each assumed to have a Hilbert space of dimension
$D$, in a quantum code \cite{Rns99}; also see p.~568 of \cite{NlCh00}.  One
argues as follows.  In order to correct up to $t$ errors on unknown carriers
the code must have a security of $s=2t$: there is no information about $\ST_a$
in any collection of $2t$ carriers, so by theorem~\ref{thm8} (ii) all the
information about $\ST_a$ is in any set of $n-2t$ carriers, as we noted
earlier.  But in a set of $n-2t$ carriers, no information can be present in a
subset of $2t$ carriers, and thus by theorem~\ref{thm8} (iii), the Hilbert
space of $n-2t-2t = n-4t$ carriers must have a dimension greater than or equal
to $d_a=K$.  This last assertion is equivalent to \eqref{eqn73}.

Note how in carrying out this argument it is essential to distinguish between a
pure state pre-probability $|\Psi\rgl$ and a mixed state pre-probability
$\rho$.  The former is needed when using theorem~\ref{thm8} (ii) to infer the
presence of all the information about $\ST_a$ in any collection of $n-2t$
carriers, given that it is absent from any collection of size $2t$.  However,
these $n-2t$ carriers along with $\ST_a$ form a system whose pre-probability is
a density operator, and as a consequence we cannot use the fact that no $2t$ of
\emph{these} carriers contain information to infer that it must be present in a
set of $n-4t$ carriers, something that is (at least in general) not true.  By
using theorem~\ref{thm8} (iii) instead of theorem~\ref{thm8} (ii), we correctly
infer that the leftover collection of $n-4t$ carriers has a certain minimal
size, \emph{not} that it contains all the information!

The foregoing discussion focussed on codes for which arbitrary errors in $t$ or
fewer carriers can be corrected.  What of codes designed for the correction of
errors of a more specific sort?  Once again \eqref{eqn71} applies, but only to
a more specialized class of operators.  Consider, for example, the three-qubit
code which is adequate for bit-flip errors, \cite{Stne98} or p.~430 of
\cite{NlCh00}.  Such errors can be represented by a Pauli $\sg^x$ on a single
qubit, and what \eqref{eqn71} is telling us is that \emph{no $\XS$ information
about any pair of qubit code carriers can be present in $\ST_a$}, where the
sample space $\XS$ is the orthonormal basis $\{|x^\pm_1\,x^\pm_2\rgl\}$ if the
carriers are 1 and 2; here $|x^\pm\rgl$ are the eigenstates of $\sg^x$.

The statement about absence at the channel input $\ST_a$ of certain types of
information about some of the carriers can be misinterpreted if thought of in
terms of some backwards-in-time ``influence'' which the carriers exert on the
channel input.  Instead, keep in mind that the real issues have to do with
statistical correlations between states-of-affairs at different times as
represented in appropriate sample spaces or frameworks.  Error recovery
depends, of course, on information being present in appropriate locations, and
quantum no-cloning (loosely speaking) allows us to connect the presence of
information in one place with its absence someplace else.  Presence and absence
should always be thought of in terms of statistical correlations.

\section{Classification of Channel and Entanglement Problems}
\label{sct6}

The fact that the properties of a quantum channel can be deduced from those of
a channel ket, and likewise the properties of an entangled mixed state from
those of a suitable purification, suggest the possibility of classifying these
two types of quantum information problem in a single scheme based on pure
states of a $p$-part system.  Of course, for each $p$ one should then introduce
additional categories with some information-theoretical significance.  The
dimensions of the $p$ subsystems are meaningful parameters, and other
features, such as the ``all'' or ``nothing'' character of certain types of
information, could assist in classifying particular cases.
The motivation behind such a classification scheme is to have a useful way of
comparing different types of experimental phenomena or theoretical models,
one that may suggest analogies in instances where these are not immediately
evident. Seeing how it relates to other problems does not, of course,
automatically provide a solution or even a better way of thinking about a
particular question, but could in some cases suggest an alternative approach,
or allow the application of a different set of ideas.

There are two reasons for preferring a classification using entangled states to
one based on channels.  First, every channel problem (of the sort under
discussion) maps in a simple and natural way to an entanglement problem, while
the reverse is subject to some qualifications, as discussed in
Sec.~\ref{sct2e}.  Second, entangled states have a higher ``conceptual
symmetry''; for example, it is more natural to ask what happens if two
subsystems of a bipartite system are interchanged than what will occur if the
channel is, so-to-speak, operated in a time-reverse mode.  The utility of pure
states as against mixed states is less obvious, but the results in
Sec.~\ref{sct4} suggest that this may lead to a simpler classification using
the location of quantum information, assuming that is a useful way to proceed.

Now let us consider some preliminary results.  The Schmidt expansion for
bipartite pure states provides a complete classification, up to local
unitaries, for $p=2$, and the by now standard pure-state entanglement measure
has proven itself a remarkably useful tool for their study.
Noiseless quantum channels described by unitary time development fall in this
category, and correspond to fully-entangled states.

The difficult problems start with $p=3$, which includes both mixed-state
entanglement and the standard model for noisy quantum channels.  Classifying
the two together immediately raises the question of how various mixed-state
entanglement measures, \cite{Kyl02,DnHR02}, may be related to the many
different types of quantum channel capacity that have been defined
\cite{Kyl02,KrWr04}.  There is a brief discussion in 
Sec.~6.3.3 of \cite{Kyl02}, which notes that the equivalence of the
(simple) quantum capacity and a one-way distillation entanglement measure was
demonstrated in \cite{Bnao96}.  But we know of no systematic attempt to relate
objects which ought to have a close connection.  Or, if they do not have
a close connection, why is that? 

If one further classifies $p=3$ problems according to the sizes of the
subsystems, the obvious starting point is pure states of three qubits. Some
one-qubit noisy channel problems fall in this category, as does the simplest
cloning problem \cite{NiGr99}.  Leaving aside cases of a product state of one
qubit with an entangled state of the other two, which in some sense belong to
the $p=2$ class, the remaining states fall into two classes, ``W'' and ``GHZ,''
under the equivalence generated by
\begin{equation}
  |\Psi'\rgl = (A\ot B\ot C)|\Psi\rgl,
\label{eqn74}
\end{equation}
where $A$, $B$, and $C$ are nonsingular operators \cite{DrVC00}.  This is a
very interesting result which does not seem to have been generalized to larger
subsystems.  However, even for qubits it may not represent a complete
classification scheme, for operations of the form \eqref{eqn74} do not, in
general, preserve all the properties that are of interest from an
information-theoretic perspective (in which $|\Psi\rgl$ functions as a
pre-probability).

The general one-qubit noisy quantum channel falls in the $p=3$ category, with
two subsystems (entrance and exit of the channel) of dimension 2, and one (the
environment) of dimension 4.  A quite general description of such channels has
been worked out in \cite{RsSW02}, and this work can and should be regarded as a
significant step in classifying a large and important set of tripartite pure
states.  There are, on the other hand, entangled states which escape this
classification (for the reasons explained in Sec.~\ref{sct2e}), and it would be
interesting if the methods used in \cite{RsSW02} could be extended to these as
well.

A unitary transformation mapping a bipartite system to itself can be thought of
as a $p=4$ problem, equivalent to a fully-entangled state between two bipartite
systems.  In the case of two qubits such unitaries can be written down
explicitly in terms of three real parameters \cite{KrCr01}, up to local
unitaries on the individual qubits, and this provides a convenient description
of an important class of $p=4$ pure states in which each subsystem has
dimension two.  Beyond this very little seems to be known at present about the
four qubit problem.  A one-qubit channel with a mixed-state environment falls
in this category, as explained in Sec.~\ref{sct5a}.  The \emph{entanglement of
purification} introduced in \cite{Trao02} is an example of a $p=4$ problem not
limited to qubits, as is the general problem of a channel corresponding to a
mixed-state environment.

As noted in Sec.~\ref{sct5c}, a quantum code with $n$ carriers falls in the
$p=n+1$ category of states for which there is an absence of correlations
between one particular subsystem (the channel entrance) and various collections
of other subsystems.  Relating quantum codes to more general problems of
multipartite entanglement is an interesting and challenging problem
\cite{Sctt04}.

\section{Conclusion}
\label{sct7}

\subsection{Summary}
\label{sct7a}

The fundamental idea underlying the duality discussed in Sec.~\ref{sct2} is
that the correlation of events at different times that characterize a quantum
channel are ``the same thing'' as the correlation of properties of an entangled
quantum system at different points in space.  At the mathematical level the
correspondence is expressed by a simple partial transpose \eqref{eqn23} that
carries the dynamical density operator $R$, into the transition operator $Q$
representing the channel superoperator.  In physical terms the duality says
that the correlations which express the location of information about one
quantum system in another are of basically the same nature, whether they refer
to properties of a single system at two different times, or to two different
systems at the same time.  This is well-established in classical information
theory, where the same tools are used for both circumstances, and it works
equally well in quantum systems given appropriate sample spaces or frameworks,
as explained in Sec.~\ref{sct3}.

The nonclassical ``peculiarities'' of quantum information emerge when one uses
a single pre-probability, either a pure state or a density operator, or their
counterparts for a quantum channel, to generate probability distributions and
thus correlations for a variety of different, incompatible frameworks (sample
spaces).  It is here that ``no-cloning'' plays a central role, and the eight
all-or-nothing theorems of Sec.~\ref{sct4} are intended to make that idea more
precise and more widely applicable.  While the theorems are expressed in
entanglement language, the duality allows their immediate application to
quantum channels.  In many cases the results are more precise (and in others
their derivation is easier) when the pre-probability is a pure rather than a
mixed state, which in the case of a quantum channel means a channel ket rather
than a dynamical operator.  This suggests that channel kets are a useful tool
for analyzing the properties of noisy quantum channels, and the applications in
Sec.~\ref{sct5} bear this out.  Whether pure states are equally advantageous
for classifying entangled states and quantum channels in a single scheme
remains to be demonstrated, but the preliminary results in Sec.~\ref{sct6} are
encouraging.

\subsection{Open questions}
\label{sct7b}

The eight all-or-nothing theorems of Sec.~\ref{sct4} provide a useful first
step in describing in a systematic way how information can be divided up or
spread out over an entangled quantum system.  But one suspects there remains
much more to be said, both about bipartite and tripartite systems, and also
about systems with $p\geq 4$ parts.
In addition, every qualitative theorem of the type found in Sec.~\ref{sct4}
ought to be the limiting case of one or perhaps several \emph{quantitative}
theorems in which the complete presence or absence of information is replaced
by quantitative measures --- Shannon entropies are an obvious, but not the
unique possibility --- and constraints are provided in the form or rigorous
inequalities, or perhaps even equalities, if one is lucky.  While some ideas of
this sort have been put forward, e.g., \cite{Hll95,Hll97}, a great deal more
could be done.

To be sure, several entanglement measures have been proposed for bipartite
mixed states~\cite{Kyl02,DnHR02}, and to a lesser extent for systems with
$p\geq 3$ parts; see \cite{Sctt04} and the references given there.  But rarely
do these have a specific information-theoretical content or basis, and it is an
open question whether, and if so how, they can be understood in such terms,
i.e., related to statistical correlations forming part of a consistent
probabilistic description of a quantum system.  To be sure, entanglement
measures can be useful even if they have no connection to information theory,
but if there is such a connection, understanding what it is could be a useful
contribution to the subject.

Discussions of quantum channel capacities seem better anchored in an
information-theoretic framework than those concerning entanglement measures,
though perhaps more thought should be given as to how to translate
``classical,'' which occurs rather frequently in such discussions, into
appropriate quantum mechanical terms; we no longer live in a classical world!
Relating these capacities to entanglement measures seems at present a largely
open question, and answering it could make a valuable contribution
understanding both entanglement and noisy channels.

The task of classifying entangled pure states of $p$-part systems in the manner
suggested in Sec.~\ref{sct6} can be regarded as complete for $p=2$, but for
$p=3$ it has just begun, and very little is known about $p\geq 4$ systems apart
from work on quantum codes.  Extending the latter to more general entangled
states could make a significant contribution to our understanding of
multipartite entanglement, which at present is quite limited.

	\section*{Acknowledgments}

I thank L. Yu for providing some of the references, and for a critical reading
of the text.  The research described here received support from the National
Science Foundation through Grant PHY-0139974.

\appendix
\numberwithin{equation}{section}
\section{Appendix. Proofs of theorems in Sec.~\ref{sct4}}
\label{scta}

Theorem~\ref{thm1} (i). Expand $|\Psi\rgl$ in Schmidt form,
\begin{equation}
   |\Psi\rgl = \sum_j \sqrt{q_j}\,|a^j\rgl\ot |c^j\rgl,
\label{Aeqn1}
\end{equation}
and let $J$ be the collection of $j$ values for which $q_j >0$.  For the 
$\{A^l\}$ information to be absent from $\ST_c$, it must be the case,
see \eqref{eqn36}, that
\begin{equation}
  \Tr_a(\Psi A^l) = \sum_{j,k}\sqrt{p_j p_k}\,
 \lgl a^j|A^l|a^k\rgl\Bigl(|c^j\rgl\lgl c^k|\Bigr)
\label{Aeqn2}
\end{equation}
is proportional to
\begin{equation}
  \Psi_c = \sum_{j\in J}p_j[c^j],
\label{Aeqn3}
\end{equation}
which means that 
\begin{equation}
  \lgl a^j|A^l|a^k\rgl = a_l\dl_{jk}
\label{Aeqn4}
\end{equation}
for all $j$ and $k$ in $J$.  This is the same as \eqref{eqn43}, which is the
same as \eqref{eqn42}.

Theorem~\ref{thm1} (ii). Expand $|\Psi\rgl$ in the orthonormal basis
$\{|a^j\rgl\}$ (see \eqref{eqn6}):
\begin{equation}
  |\Psi\rgl = \sum_j |a^j\rgl\ot|\gm^j\rgl.
\label{Aeqn5}
\end{equation}
The requirement that no information about $\{|a^j\rgl\}$ be in $\ST_c$ means
that all the $|\gm^j\rgl$ must be proportional to each other, and thus to a
single ket $|\gm\rgl$, which means that $|\Psi\rgl$ is of the form
\eqref{eqn44}.

Theorem~\ref{thm1} (iii). The ``if'' part is obvious.
To prove that \eqref{eqn45} holds if all information about $\ST_a$ is absent
from $\ST_c$, let $\{|a^j\rgl\}$ and $\{|c^l\rgl\}$
be bases in which $\rho_a$ and $\rho_c$ are diagonal, 
\begin{equation}
  \rho_a=\sum_j p_j[a^j],\quad \rho_c = \sum_l q_l [c^l],
\label{Aeqn6}
\end{equation}
and write
\begin{equation}
 \rho= \sum_{jk}\sum_{lm} \lgl a^j c^l|\rho|a^{k} c^{m}\rgl 
 \Bigl(|a^j\rgl\lgl a^{k}|\ot |c^l\rgl\lgl c^{m}|\Bigr)
\label{Aeqn7}
\end{equation}
The absence of all information implies that 
\begin{equation}
  \Tr_a(A\rho) = \lgl A\rgl \rho_c
\label{Aeqn8}
\end{equation}
for \emph{any} operator $A\,\in\hat\HS_a$ --- see \eqref{eqn36}, and note that
the collection of all projectors is an operator basis for $\hat\HS_a$.  Insert
$A=|a^{k}\rgl\lgl a^j|$ in \eqref{Aeqn8}, and use \eqref{Aeqn7} to evaluate
the left side and \eqref{Aeqn6} the right.  The conclusion is that
\begin{equation}
   \lgl a^j c^l|\rho|a^{k} c^{m}\rgl = p_j q_l \dl_{jk} \dl_{lm},
\label{Aeqn9}
\end{equation}
which is \eqref{eqn45}.

Theorem~\ref{thm2} (i). If the $\AS$ information is present in $\ST_b$, 
\eqref{eqn35} implies that
\begin{equation}
  \Tr(A^l B^m)=\Tr_b(\Lm^l B^m) = \dl_{lm}\Tr_b(\Lm^l),
\label{Aeqn10}
\end{equation}
since $\lgl A^l\rgl=\Tr_b(\Lm^l)$.  If $P$ and $Q$ are positive operators such
that $\Tr(PQ)=0$, then $PQ=0$.  Using this and the fact that the $B^m$ are
projectors, so that $\Lm^l=\Lm^l B^m + \Lm^l(I_b- B^m)$, one sees that
\eqref{Aeqn10} implies that
\begin{equation}
  B^m \Lm^l B^m = \dl_{lm}\Lm^l, 
\label{Aeqn11}
\end{equation}
and \eqref{eqn46} is a consequence of $B^l B^m = \dl_{lm}B^l$.  Conversely,
\eqref{eqn46} implies that one can simultaneously diagonalize the collection
$\{\Lm^l\}$ and choose the $B^l$ projecting onto appropriate blocks in such a
way that \eqref{Aeqn11}, and therefore \eqref{Aeqn10} and \eqref{eqn35} are
satisfied.

Theorem~\ref{thm2} (ii). Choose an orthonormal basis $\{|a^j\rgl\}$ in which the
$\{A^l\}$ are diagonal, and expand $|\Psi\rgl$ in this basis, \eqref{eqn49},
without assuming it is in Schmidt form. Then
\begin{equation}
  \Psi_a = \sum_{jk}\lgl\bt^k|\bt^j\rgl\; |a^j\rgl\lgl a^k|
\label{Aeqn12}
\end{equation}
and 
\begin{equation}
  \Lm^l = \sum_{j\in J_l} |\bt^j\rgl\lgl\bt^j|,
\label{Aeqn13}
\end{equation}
where $J_l$ is the collection of $j$ values for which $A^l |a^j\rgl =|a^j\rgl$.
One can show that $\Psi_a$ commutes with all the $A^l$ if and only if
$\lgl\bt^k|\bt^j\rgl=0$ whenever $j\in J_l$ and $k\in J_m$ with $m\neq l$.  But
this last is equivalent to \eqref{eqn46}.  If the $A^l$ project onto
one-dimensional states, then $\lgl\bt^k|\bt^j\rgl=0$ for $j\neq k$,
so \eqref{eqn49} is in Schmidt form.

Theorem~\ref{thm2} (iii). Purify $\rho$ to a ket $|\Psi\rgl\,\in\HS_{abc}$. Use
the fact that the $\AS$ information is present in $\ST_{bc}$, and apply part
(ii) of the theorem with $\ST_{bc}$ in place of $\ST_b$ to infer that
$\Psi_a=\rho_a$ commutes with all the $A^l$.

Theorem~\ref{thm3}. Part (i) is an immediate consequence of 2(ii), for it is
only multiples of the identity that commute with all projectors.  The proofs of
(ii) and (iii) are given below, following that of theorem~\ref{thm8}.

Theorem~\ref{thm4}.  By theorem~\ref{thm2}, $\rho_a$ or $\Psi_a$ must commute
with all the $\{A^j\}$ and all the $\{\bar A^k\}$, and must therefore, by the
definition of strong incompatibility, be multiples of $I_a$.  The final
statement is a consequence of theorem~\ref{thm3}

Theorem~\ref{thm5}. Let $\{|a^j\rgl\}$ and $\{|c^k\rgl\}$ be orthonormal bases
of $\HS_a$ and $\HS_c$ which diagonalize $\Psi_a$ and $\Psi_c$,
\begin{equation}
  \Psi_a = \sum_j p_j [a^j],\quad \Psi_b = \sum_k q_k [c^k],
\label{Aeqn14}
\end{equation}
and expand $|\Psi\rgl$ in these bases:
\begin{equation}
  |\Psi\rgl = \sum_{jk} |a^j\rgl\ot|\bt^{jk}\rgl \ot |c^k\rgl.
\label{Aeqn15}
\end{equation}
The condition $\Psi = \Psi_a\ot\Psi_c$ expressing the absence of all $\ST_a$
information from $\ST_b$, theorem~\ref{thm1} (iii), implies that
\begin{equation}
  \lgl\bt^{j'k'}|\bt^{jk}\rgl = p_j q_k\dl_{jj'}\dl_{kk'}.
\label{Aeqn16}
\end{equation}
Therefore if we restrict our attention to the $j\in J$ and $k\in K$ for which
$p_j>0$ and $q_k >0$, we can construct an orthonormal set
\begin{equation}
  |b^{jk}\rgl = |\bt^{jk}\rgl/\sqrt{p_j q_k}
\label{Aeqn17}
\end{equation}
of kets in $\HS_b$, and rewrite \eqref{Aeqn15} in the form
\begin{equation}
 |\Psi\rgl =  \sum_{jk} \sqrt{p_j q_k}\, |a^j\rgl\ot|b^{jk}\rgl \ot |c^k\rgl.
\label{Aeqn18}
\end{equation}
The spaces $\HS_d$ and $\HS_e$ are then \emph{defined} as having orthonormal
bases $\{|d^j\rgl\}$ and  $\{|e^k\rgl\}$ such that
\begin{equation}
  |b^{jk}\rgl = |d^j\rgl\ot |e^k\rgl,
\label{Aeqn19}
\end{equation}
so that $|\Psi\rgl$ is of the form \eqref{eqn56} with 
\begin{equation}
  |\chi\rgl = \sum_j \sqrt{p_j}\, |a^j\rgl\ot |d^j\rgl,\quad 
  |\psi\rgl = \sum_k \sqrt{q_k}\, |c^k\rgl\ot |e^k\rgl.
\label{Aeqn20}
\end{equation}

Theorem~\ref{thm6} (i).  Expand $|\Psi\rgl$ in the orthonormal
basis $\{|a^j\rgl\}$
\begin{equation}
  |\Psi\rgl = \sum_j |a^j\rgl\ot|\zt^j\rgl,
\label{Aeqn21}
\end{equation}
with $|\zt^j\rgl\in\HS_{bc}$, and write
\begin{equation}
  \Psi = \sum_{jk} |a^j\rgl\lgl a^k|\ot \zt^{jk};\quad 
  \zt^{jk} := |\zt^j\rgl\lgl\zt^k|.
\label{Aeqn22}
\end{equation}
If the $\{|a^j\rgl\}$ information is in $\ST_b$, then by theorem~\ref{thm2} (i)
\begin{equation}
  \zt^{jj}_b \zt^{kk}_b = 0 \text{ for } j\neq k,
\label{Aeqn23}
\end{equation}
where, following our usual notation, $\zt^{jj}_b=\Tr_c(\zt^{jj})$. Now apply 
\eqref{Beqn3} in App.~\ref{sctb}, with $a$ replaced by $b$, $b$ replaced by
$c$, $|e\rgl=|\zt^j\rgl$ and 
$|g\rgl=|\zt^k\rgl$, to conclude that \eqref{Aeqn23} holds if and only if
\begin{equation}
  \zt^{jk}_c = 0  \text{ for } j\neq k.
\label{Aeqn24}
\end{equation}
(Note that $\Tr_c(CC\ad)=0$ implies that $C=0$.) But \eqref{Aeqn24} inserted in
\eqref{Aeqn22} implies \eqref{eqn57} with $\Gm^j = \zt^{jj}_c$. Conversely,
\eqref{eqn57} implies \eqref{Aeqn24}, which implies \eqref{Aeqn23}, which,
using theorem~\ref{thm2} (i), implies that the $\{|a^j\rgl\}$ information is in
$\ST_b$.

Theorem~\ref{thm6} (ii).  Let $\{|a^j\rgl\}$ be any basis in which the $A^k$ in
\eqref{eqn58} are diagonal.  Then \eqref{eqn57} is a consequence of
\eqref{eqn58}: simply write each $A^k$ as a sum of a suitable collection of
$[a^j]$. Thus by (i), the $\{|a^j\rgl\}$ and, a fortiori the $\{A^k\}$
information is in $\ST_b$.  For a compatible decomposition $\bar\AS=\{\bar
A^l\}$, use a basis $\{|a^j\rgl\}$ in which both these and the $\{A^k\}$
are diagonal.

Theorem~\ref{thm7} (i). Purify $\rho$ to $|\Psi\rgl\in\HS_{abcd}$, and apply
theorem~\ref{thm6} (i) with $c$ replaced by $cd$ to conclude that $\Psi_{acd}$
is of the form \eqref{eqn57}  with operators $\Gm^j$ on $\HS_{cd}$.  Now trace
both sides over $\HS_d$ to get the equivalent of \eqref{eqn59}.  

Theorem~\ref{thm7} (ii). Multiply both sides of \eqref{eqn59} by $[\bar a^k]$. 
First trace over $\HS_a$ and use the definition of mutually unbiased bases
in \eqref{eqn4} to conclude that the resulting operator (on $\HS_c$) does not
depend on $k$, so the $\{|\bar a^k\rgl\}$ information is absent from $\ST_c$
according to the definition in Sec.~\ref{sct3c}, see the comment following
\eqref{eqn36}.  Next, trace over $\HS_c$ to get \eqref{eqn60}.

Theorem~\ref{thm8} (i).  Given an arbitrary orthonormal basis $\bar\AS$ of
$\HS_a$, one can always find another basis $\AS$ with $\bar\AS$ and $\AS$
mutually unbiased.  As the $\AS$ information is, by assumption, in $\ST_b$, 
the $\bar\AS$ information cannot be in $\ST_c$, by theorem~\ref{thm7} (ii).

Theorem~\ref{thm8} (ii).  All the information about $\ST_a$ is in $\ST_{bc}$,
so $\Psi_a=I_a/d_a$ by theorem~\ref{thm3} (i).  But as there is no
information about $\ST_a$ in $\ST_c$, theorem~\ref{thm5} tells us $|\Psi\rgl$
is of the form \eqref{eqn56}, with $\chi_a=\Psi_a=I_a/d_a$, and therefore, once
again invoking theorem~\ref{thm3} (i), all the information about $\ST_a$ is in
$\ST_b$.

Theorem~\ref{thm8} (iii). (The following argument is from p.~569 of
\cite{NlCh00}, where it is ascribed to \cite{Prsk98}, and it makes use of some
well-known properties of the von Neumann entropy
\begin{equation}
  S(\rho) = -\Tr(\rho\log\rho);
\label{Aeqn25}
\end{equation}
see, e.g., pp.~513 and 515 of
\cite{NlCh00}.)
Upon purifying $\rho$ to  $|\Psi\rgl\in\HS_{abcd}$ one finds that
\begin{equation}
  S(\Psi_a)+S(\Psi_c) = S(\Psi_{ac}) = S(\Psi_{bd}) \leq S(\Psi_b)+S(\Psi_d).
\label{Aeqn26}
\end{equation}
The first equality is a consequence of the absence of information about $\ST_a$
in $\ST_c$, thus $\Psi_{ac}=\Psi_a\ot\Psi_c$ by theorem~\ref{thm1} (iii). The
second equality reflects the fact that $|\Psi\rgl$ is a pure state on
$\HS_{ac}\ot\HS_{bd}$, and the final inequality is a standard result for a
density operator on a tensor product.  Since all information about $\ST_a$ is
in $\ST_{bc}$, it must be absent from $\ST_d$ by part (i) of this theorem, so
we can interchange the roles of $\ST_c$ and $\ST_d$ in \eqref{Aeqn26} to obtain
\begin{equation}
  S(\Psi_a)+S(\Psi_d) \leq S(\Psi_b)+S(\Psi_c),
\label{Aeqn27}
\end{equation}
and by adding this to \eqref{Aeqn26} arrive at
\begin{equation}
  S(\Psi_a)\leq S(\Psi_b).
\label{Aeqn28}
\end{equation}
By theorem~\ref{thm3} (i) (replace $b$ by $bcd$) we know that $\Psi_a=I_a/d_a$,
so the left side of \eqref{Aeqn28} is $\log d_a$, and as the right side cannot
exceed $\log d_b$, therefore $d_b\geq d_a$. 

Theorem~\ref{thm3} (ii) and (iii).  Purify $\rho$ to $|\Psi\rgl\in\HS_{abc}$.
If all information about $\ST_a$ is in $\ST_b$ (for $\rho$ and for
$|\Psi\rgl$), then by theorem~\ref{thm8} (i) there is none in $\ST_c$, so by
theorem~\ref{thm5} $|\Psi\rgl$ has the product structure of \eqref{eqn56},
where in addition $|\chi\rgl$ must be maximally (fully) entangled, so we arrive
at \eqref{eqn52}.  If, on the other hand, \eqref{eqn52} is correct, then
$\phi_a = I_a/d_a$, and all the information about $\ST_a$ is in $\ST_d$, and
therefore in $\ST_b$.  To prove theorem~\ref{thm3} (iii), note that if
$|\Psi\rgl$ is given by \eqref{eqn56} and $d_e$ is 2 or more, $I_e$ has a
nontrivial decomposition, and the corresponding information obviously cannot be
in $\ST_a$.  Thus if all the information about $\ST_b$ is in $\ST_a$, it is the
case that $d_e=1$ and $\HS_{de}$ is the same as $\HS_d$, and the latter is the
same as $\HS_b$, for were it a proper subspace, $\Psi_b$ would not be
proportional to $I_b$.

\section{Appendix. Four entangled kets}
\label{sctb}


Let 
\begin{equation}
  D^{ef} = |e\rgl\lgl f|,\quad D^{ef}_a = \Tr_b(D^{ef}),\quad
  \quad D^{ef}_b = \Tr_a(D^{ef})
\label{Beqn1}
\end{equation}
denote the dyad and its partial traces for two kets $|e\rgl$ and $|f\rgl$
on $\HS_{ab}=\HS_a\ot\HS_b$. 

Theorem. Let $|e\rgl$, $|f\rgl$, $|g\rgl$, $|h\rgl$ be any four kets
on $\HS_{ab}$. Then
\begin{equation}
  \Tr_a(D^{ef}_a D^{gh}_a ) = \Tr_b(D^{eh}_b D^{gf}_b ).
\label{Beqn2}
\end{equation}
In particular, if $|f\rgl = |e\rgl$ and $|h\rgl = |g\rgl$, then
\begin{equation}
  \Tr_a(D^{ee}_a D^{gg}_a ) = \Tr_b(D^{eg}_b D^{ge}_b ).
\label{Beqn3}
\end{equation}

Proof. Let $\{|a^j\rgl\}$ be a fixed orthonormal basis of $\HS_a$, and expand
each ket in the form
\begin{equation}
  |w\rgl = \sum_j |a^j\rgl\ot |w^j\rgl.
\label{Beqn4}
\end{equation}
Direct calculation shows that the left and right sides of \eqref{Beqn2} are
both equal to
\begin{equation}
  \sum_{jk} \lgl f^k|e^j\rgl\lgl h^j|g^k\rgl.
\label{Beqn5}
\end{equation}


\begin{thebibliography}{10}

\bibitem{Kyl02}
Michael Keyl.
\newblock Fundamentals of quantum information theory.
\newblock {\em Phys. Rep.}, 369:431--548, 2002.

\bibitem{VrVr03}
Frank Verstraete and Henri Verschelde.
\newblock On quantum channels.
\newblock quant-ph/0202124, 2003.

\bibitem{ZcBn04}
Karol \.Zyczkowski and Ingemar Bengtsson.
\newblock On duality between quantum maps and quantum states.
\newblock {\em Open Syst. Inf. Dyn.}, 3:42, 2004.

\bibitem{ArPt04}
Pablo Arrighi and Christophe Patricot.
\newblock On quantum operations as quantum states.
\newblock {\em Ann. Phys. (NY)}, 311:26--52, 2004.

\bibitem{CDKL01}
J.~I. Cirac, W.~D{\"u}r, B.~Kraus, and M.~Lewenstein.
\newblock Entangling operations and their implementation using a small amount
  of entanglement.
\newblock {\em Phys. Rev. Lett.}, 86:544--547, 2001.

\bibitem{RmWr03}
M.~Reimpell and R.~F. Werner.
\newblock Iterative optimization of quantum error correcting codes.
\newblock quant-ph/0307138, 2003.

\bibitem{HrSR03}
Michael Horodecki, Peter~W. Shor, and Mary~Beth Ruskai.
\newblock Entanglement breaking channels.
\newblock {\em Rev. Math. Phys.}, 15:629--641, 2003.
\newblock quant-ph/0302031.

\bibitem{Rska03}
Mary~Beth Ruskai.
\newblock Entanglement breaking channels.
\newblock {\em Rev. Math. Phys.}, 15:643--662, 2003.
\newblock quant-ph/0302032.

\bibitem{Hmda03}
Mitsuru Hamada.
\newblock Notes on the fidelity of symplectic quantum error-correcting codes.
\newblock {\em Int. J. Quantum Inf.}, 1:443--463, 2003.

\bibitem{DCHr04}
W.~D{\"u}r, J.~I. Cirac, and P.~Horodecki.
\newblock Nonadditivity of quantum capacity for multiparty communication
  channels.
\newblock {\em Phys. Rev. Lett.}, 93:020503, 2004.

\bibitem{Jmlk72}
A.~{Jamio\l kowski}.
\newblock Linear transformations which preserve trace and positive
  semidefiniteness of operators.
\newblock {\em Reports on Mathematical Physics}, 3:275--278, 1972.

\bibitem{Choi75}
Man-Duen Choi.
\newblock Completely positive linear maps on complex matrices.
\newblock {\em Lin. Alg. Appl.}, 10:285--290, 1975.

\bibitem{BrZl99}
{\u C}aslav Brukner and Anton Zeilinger.
\newblock Operationally invariant information in quantum measurements.
\newblock {\em Phys. Rev. Lett.}, 83:3354--3357, 1999.

\bibitem{BrZl01}
{\u C}aslav Brukner and Anton Zeilinger.
\newblock Conceptual inadequacy of the {S}hannon information in quantum
  measurements.
\newblock {\em Phys. Rev. A}, 63:022113, 2001.

\bibitem{DtHy00}
David Deutsch and Patrick Hayden.
\newblock Information flow in entangled quantum systems.
\newblock {\em Proc. R. Soc. London A}, 456:1759--1774, 2000.

\bibitem{Grff02}
Robert~B. Griffiths.
\newblock Nature and location of quantum information.
\newblock {\em Phys. Rev. A}, 66:012311, 2002.

\bibitem{Tmps03}
C.~G. Timpson.
\newblock On a supposed conceptual inadequacy of the {S}hannon information in
  quantum mechanics.
\newblock {\em Stud. Hist. Phil. Mod. Phys.}, 34:441--468, 2003.

\bibitem{Dwll03}
Armond Duwell.
\newblock Quantum information does not exist.
\newblock {\em Stud. Hist. Phil. Mod. Phys.}, 34:479--499, 2003.

\bibitem{NlCh00}
Michael~A. Nielsen and Isaac~L. Chuang.
\newblock {\em Quantum Computation and Quantum Information}.
\newblock Cambridge University Press, Cambridge, 2000.

\bibitem{Sctt04}
A.~J. Scott.
\newblock Multipartite entanglement, quantum-error-correcting codes, and
  entangling power of quantum evolutions.
\newblock {\em Phys. Rev. A}, 69:052330, 2004.

\bibitem{Fllr68}
William Feller.
\newblock {\em An introduction to probability theory and its applications},
  volume~1.
\newblock Wiley, New York, 3d edition, 1968.

\bibitem{Rss00}
Sheldon~M. Ross.
\newblock {\em Introduction to probability models}.
\newblock Academic Press, San Diego, 7th edition, 2000.

\bibitem{DGSc02}
Morris~H. DeGroot and Mark~J. Schervish.
\newblock {\em Probability and Statistics}.
\newblock Addison-Wesley, Boston, 3d edition, 2002.

\bibitem{Grff02b}
Robert~B. Griffiths.
\newblock {\em Consistent Quantum Theory}.
\newblock Cambridge University Press, Cambridge, U.K., 2002.

\bibitem{Mttl98}
Peter Mittelstaedt.
\newblock {\em The Interpretation of Quantum Mechanics and the Measurement
  Process}.
\newblock Cambridge, Cambridge, U.K., 1998.

\bibitem{dMyn02}
Willem~M. de~Muynck.
\newblock {\em Foundations of Quantum Mechanics, an Empiricist Approach}.
\newblock Kluwer Academic Publishers, Dordrecht, The Netherlands, 2002.

\bibitem{Trao99}
Barbara~M. Terhal, Isaac~L. Chuang, David~P. DiVincenzo, Markus Grassl, and
  John~A. Smolin.
\newblock Simulating quantum operations with mixed environments.
\newblock {\em Phys. Rev. A}, 60:881--885, 1999.

\bibitem{ZlRf02}
Christof Zalka and Eleanor Rieffel.
\newblock Quantum operations that cannot be implemented using a small mixed
  environment.
\newblock {\em J. Math. Phys.}, 43:4376--4381, 2002.

\bibitem{Hlvo99}
Alexander~S. Holevo.
\newblock Coding theorems for quantum channels.
\newblock {\em Russian Math. Surveys}, 53:1295--1331, 1999.

\bibitem{Shr02}
Peter~W. Shor.
\newblock Additivity of the classical capacity of entanglement-breaking quantum
  channels.
\newblock {\em J. Math. Phys.}, 43:4334--4340, 2002.

\bibitem{Lwao00}
M.~Lewenstein, D.~Bru\ss, J.~I. Cirac, B.~Kraus, M.~Ku\'s, J.~Samsonowicz,
  A.~Sanpera, and R.~Tarrach.
\newblock Separability and distillability in composite quantum systems -a
  primer-.
\newblock {\em J. Mod. Optics}, 47:2841, 2000.

\bibitem{HrHH01}
Micha\l\ Horodecki, Pawe\l\ Horodecki, and Ryszard Horodecki.
\newblock Separability of $n$-particle mixed states: necessary and sufficient
  conditions in terms of linear maps.
\newblock {\em Phys. Lett. A}, 283:1--7, 2001.

\bibitem{Stne98}
Andrew~W. Steane.
\newblock Quantum error correction.
\newblock In Hoi-Kwong Lo, Sandu Popescu, and Tim Spiller, editors, {\em
  Introduction to Quantum Computation and Information}, pages 184--212. World
  Scientific, Singapore, 1998.

\bibitem{Bnao96}
Charles~H. Bennett, David~P. DiVincenzo, John~A. Smolin, and William~K.
  Wootters.
\newblock Mixed-state entanglement and quantum error correction.
\newblock {\em Phys. Rev. A}, 54:3824--3851, 1996.

\bibitem{KnLf97}
Emanuel Knill and Raymond Laflamme.
\newblock Theory of quantum error-correcting codes.
\newblock {\em Phys. Rev. A}, 55:900--911, 1997.

\bibitem{LMPZ96}
Raymond Laflamme, Cesar Miquel, Juan~Pablo Paz, and Wojciech~Hubert Zurek.
\newblock Perfect quantum error correcting code.
\newblock {\em Phys. Rev. Lett.}, 77:198--201, 1996.

\bibitem{GrBP97}
M.~Grassl, Th. Beth, and T.~Pellizzari.
\newblock Codes for the quantum erasure channel.
\newblock {\em Phys. Rev. A}, 56:33--38, 1997.

\bibitem{Rns99}
Eric~M. Rains.
\newblock Nonbinary quantum codes.
\newblock {\em IEEE Trans. Inf. Theory}, 45:1827--1832, 1999.

\bibitem{DnHR02}
Matthew~J. Donald, Michal Horodecki, and Oliver Rudolph.
\newblock The uniqueness theorem for entanglement measures.
\newblock {\em J. Math. Phys.}, 43:4252--4272, 2002.

\bibitem{KrWr04}
Dennis Kretschmann and Reinhard~F Werner.
\newblock Tema con variazioni: quantum channel capacity.
\newblock {\em New J. Phys.}, 6:26, 2004.

\bibitem{NiGr99}
Chi-Sheng Niu and Robert~B. Griffiths.
\newblock Two-qubit copying machine for economical quantum eavesdropping.
\newblock {\em Phys. Rev. A}, 60:2764--2776, 1999.

\bibitem{DrVC00}
W.~D{\"u}r, G.~Vidal, and J.~I. Cirac.
\newblock Three qubits can be entangled in two inequivalent ways.
\newblock {\em Phys. Rev. A}, 62:062314, 2000.

\bibitem{RsSW02}
Mary~Beth Ruskai, Stanislaw Szarek, and Elisabeth Werner.
\newblock An analysis of completely-positive trace-preserving maps on
  $2\times2$ matrices.
\newblock {\em Linear Algebr. Appl.}, 347:159--187, 2002.
\newblock quant-ph/0101003.

\bibitem{KrCr01}
B.~Kraus and J.~I. Cirac.
\newblock Optimal creation of entanglement using a two-qubit gate.
\newblock {\em Phys. Rev. A}, 63:062309, 2001.

\bibitem{Trao02}
Barbara~M. Terhal, Michal Horodecki, Debbie~W. Leung, and David~P. DiVincenzo.
\newblock The entanglement of purification.
\newblock {\em J. Math. Phys.}, 43:4286--4298, 2002.

\bibitem{Hll95}
Michael J.~W. Hall.
\newblock Information exclusion principle for complementary observables.
\newblock {\em Phys. Rev. Lett.}, 74:3307--3311, 1995.

\bibitem{Hll97}
Michael J.~W. Hall.
\newblock Quantum information and correlation bounds.
\newblock {\em Phys. Rev. A}, 55:100--113, 1997.

\bibitem{Prsk98}
John Preskill.
\newblock Lecture notes, 1998.
\newblock www.theory.caltech.edu/people/preskill/ph229/.

\end{thebibliography}

\end{document}